\documentclass[a4paper,fleqn,usenatbib]{mnras}

\usepackage{newtxtext,newtxmath}

\usepackage[T1]{fontenc}
\usepackage{ae,aecompl}

\usepackage{multirow}
\usepackage{wasysym}
\usepackage{mathtools}

\usepackage{graphicx}    
\usepackage{amsmath}    
\usepackage{amssymb}    
\usepackage{todonotes}

\usepackage{array}

\newcolumntype{x}[1]{>{\centering\arraybackslash\hspace{0pt}}p{#1}}

\title[Holistic spectroscopy]{Holistic spectroscopy: Complete reconstruction of a wide-field, multi-object spectroscopic image using a photonic comb}

\author[J. Kos et al.]{
Janez Kos,$^{1}$\thanks{E-mail: janez.kos@sydney.edu.au}
Joss Bland-Hawthorn,$^{1,2,3}$
Christopher H. Betters,$^{1,3}$\newauthor
Sergio Leon-Saval,$^{1,3}$
Martin Asplund,$^{4}$
Sven Buder,$^{5}$
Andrew R. Casey,$^{6,7}$\newauthor
Valentina D'Orazi,$^{8}$
Gayandhi de Silva,$^{1,9}$
Ken Freeman,$^{4}$
Geraint Lewis,$^{1}$
Jane Lin,$^{4}$\newauthor
Sarah L. Martell,$^{10}$
Katharine Schlesinger,$^{4}$
Sanjib Sharma,$^{1}$\newauthor
Jeffrey D. Simpson,$^{9}$
Daniel Zucker,$^{9,11,12}$
Toma\v{z} Zwitter, $^{13}$\newauthor
Michael Hayden,$^{1,2}$
David M. Nataf,$^{14}$
and Yuan-Sen Ting$^{15,16,17}$
\\
$^{1}$Sydney Institute for Astronomy, School of Physics, A28, The University of Sydney, NSW 2006, Australia\\
$^{2}$ARC Centre of Excellence for All Sky Astrophysics in 3 Dimensions (ASTRO-3D)\\
$^{3}$Sydney Astrophotonic Instrumentation Labs, School of Physics, A28, The University of Sydney, NSW 2006, Australia\\
$^{4}$Research School of Astronomy and Astrophysics, Australian National University, Cotter Road, Canberra, ACT 72611, Australia\\
$^{5}$Max-Planck-Institut for Astronomy, K\"onigstuhl 17, D-69117 Heidelberg, Germany\\
$^{6}$School of Physics and Astronomy, Monash University, Victoria, Australia\\
$^{7}$Faculty of Information Technology, Monash University, Victoria, Australia, Wellington Road, Clayton, VIC 3800, Australia\\
$^{8}$Istituto Nazionale di Astrofisica, Osservatorio Astronomico di Padova, vicolo dell'Osservatorio 5, 35122, Padova, Italy\\
$^{9}$Australian Astronomical Observatory, 105 Delhi Rd, North Ryde, NSW 2113, Australia\\
$^{10}$School of Physics, University of New South Wales, Sydney, NSW 2052, Australia\\
$^{11}$Department of Physics and Astronomy, Macquarie University, Sydney, NSW 2109, Australia\\
$^{12}$Research Centre in Astronomy, Astrophysics \& Astrophotonics, Macquarie University, Sydney, NSW 2109, Australia\\
$^{13}$Faculty of mathematics and physics, University of Ljubljana, Jadranska 19, 1000 Ljubljana, Slovenia\\
$^{14}$Center for Astrophysical Sciences and Department of Physics and Astronomy, The Johns Hopkins University,\\Baltimore, MD 21218, USA\\
$^{15}$Institute for Advanced Study, Princeton, NJ 08540, USA\\
$^{16}$Department of Astrophysical Sciences, Princeton University, Princeton, NJ 08544, USA\\
$^{17}$Observatories of the Carnegie Institution of Washington, 813 Santa Barbara Street, Pasadena, CA 91101, USA\\
}

\date{Accepted XXX. Received YYY; in original form ZZZ}

\pubyear{2018}

\begin{document}
\label{firstpage}
\pagerange{\pageref{firstpage}--\pageref{lastpage}}
\maketitle

\begin{abstract}
The primary goal of Galactic archaeology is to learn about the origin of the Milky Way from the detailed chemistry and kinematics of millions of stars. Wide-field multi-fibre spectrographs are increasingly used to obtain spectral information for huge samples of stars. Some surveys (e.g. GALAH) are attempting to measure up to 30 separate elements per star. Stellar abundance spectroscopy is a subtle art that requires a very high degree of spectral uniformity across each of the fibres. However wide-field spectrographs are notoriously non-uniform due to the fast output optics necessary to image many fibre outputs onto the detector. We show that precise spectroscopy is possible with such instruments across all fibres by employing a photonic comb -- a device that produces uniformly spaced spots of light on the CCD to precisely map complex aberrations. Aberrations are parametrized by a set of orthogonal moments with $\sim100$ independent parameters. We then reproduce the observed image by convolving high resolution spectral templates with measured aberrations as opposed to extracting the spectra from the observed image. Such a forward modeling approach also trivializes some spectroscopic reduction problems like fibre cross-talk, and reliably extracts spectra with a resolution $\sim2.3$ times above the nominal resolution of the instrument. Our rigorous treatment of optical aberrations also encourages a less conservative spectrograph design in the future.
\end{abstract}




\section{Introduction}

Since the 1980s, wide-field multi-fibre spectrographs have had a major impact on astronomy. They were originally envisaged as instruments for obtaining redshifts of distant galaxies and quasars \citep[e.g., 2dF survey,][]{colless01} although with some improvements, useful spectrophotometry became possible \citep[e.g., SDSS,][]{abazajian09}. After 2000, the emergence of Galactic archaeology led to the idea that this technology could be exploited to obtain kinematic and abundance data for thousands of stars \citep{freeman02}. Early surveys \citep[e.g., RAVE,][]{steinmetz06} were successful in obtaining radial velocities and some abundance measurements (e.g., [Fe/H], [$\alpha$/Fe]) for 500,000 stars at a resolving power of R=7500. More recent surveys like APOGEE \citep{majewski17} or GALAH \citep{desilva15} have pushed for higher resolution spectroscopy at R=22,500 and R=28,000 respectively. 

The demand for higher spectroscopic resolution has come at a cost. Packing the outputs of many fibres onto large CCDs requires fast output optics. Larger off-axis angles at the detector lead to more pronounced and more complex optical aberrations. This is well known to the wide-field imaging community who write elaborate software to correct the aberrations as a function of the location in the field \citep{stetson87}. These aberrations are measured from point sources in the field. In recent years the line between image reduction and analysis has been further blended by replacing linear procedures with iterative ones and with forward modeling \citep{brewer13}.

Until now, there has been no analogy for this type of correction in wide-field multi-object spectroscopy for a variety of reasons:
\begin{itemize}
\item The point spread function (PSF\footnote{By PSF we mean the response of the fibre assembly and the spectrograph to a monochromatic source of light. We reserve the term line spread function (LSF) for the PSF mapped into a one dimensional spectrum.}) of a spectrograph is hard to measure precisely, as it depends on the wavelength as well.
\item Optimal extraction \citep{horne86} has served us well in the past, as spectra can be produced in a controlled manner where they do not overlap much.
\item A spectrum occupies a much larger portion of the CCD than a point source and has many more free parameters (compared to only flux for a point source). Computational complexity is therefore much higher.
\item Quantity over quality has been a motivation for large surveys, so the high precision spectroscopy remained in the domain of more narrowly focused studies.
\end{itemize}
A case for more precise spectroscopy is made by highly multiplexed surveys where the main goal is chemical tagging, like GALAH. A precise and reliable measurement of abundances of individual elements in individual stars -- which usually means precisely measuring individual spectral lines -- is extremely important, because combining lesser quality measurements is rarely an option. Measurements of elemental abundances have advanced greatly from simply measuring equivalent widths. Now the shape of a large portion of the spectrum is used to determine a single abundance \citep[e.g.][]{ness15}.

Observed spectra are two dimensional objects and mapping only the resolution or the line spread function (LSF) in a reduced 1D spectrum is a simplification that does not conserve the information of the original 2D PSF. In a pursuit of high-precision spectroscopy we have to think about aberrations arising from a two-dimensional nature of astronomical observations. We attempt to replace a traditional spectral extraction with a two-dimensional forward modeling schema that can take the full 2D PSF into the account. Instead of analysing the observed image, template spectra can be used to forward model an image that will resemble the observed image as well as possible. With forward modeling the aberrations that are notoriously hard to correct in a traditional spectral extraction, like fibre cross talk \citep{sharp10}, can be accurately addressed. Such approach also permits the design of the spectrograph to be less conservative, as more fibre cross-talk and larger aberrations can be properly taken into the account.

The first objective when forward modeling a 2D spectrum is to precisely measure the PSF. We utilized a photonic comb, a device able to produce regularly spaced peaks in the frequency space with a well known shape \citep{betters16, jbh17}. A photonic comb spectrum has many advantages over an arc lamp spectrum, where lines are blended and at irregular intervals that leave regions of the wavelength space without any strong lines, which  are required for precise PSF measurements. Our approach provides an unmatched precision, offering as many as $\sim100$ independent parameters (two dimensional orthogonal moments) to describe the PSF anywhere on the CCD plane. Even more, the PSF can be produced with different fibre configurations without probing the different configurations individually.

With the PSF precisely measured and parametrized we can use synthetic template spectra and reconstruct the observed 2D image. A grid of spectra can be used in the same way as with a  single spectra to find the best matching template for each observed spectrum. We demonstrate on Solar spectra that the forward modeling approach is reliable, the reproduced image resembles the observed image in great detail, and the same template gives equally good reproduction regardless which fibre is selected and how distorted the PSF is in that point. Our approach is a practical way of an idea presented in \citet{bolton10} that can be used for any kind of multi-fibre spectroscopy, if the template matching used here for Solar spectra is replaced with a more elaborate, perhaps iterative method for production of one dimensional empirical spectra.

\section{Data acquisition}

\subsection{The 2dF and Hermes Instruments}

We demonstrate our method on spectra taken by the Hermes spectrograph. Hermes is a multi-fibre, four band spectrograph at the 3.9~m AAT telescope at the Siding Spring Observatory  \citep{sheinis15}. It covers 1000~{\AA} split into four bands between blue and NIR (4720 to 4900, 5650 to 5880, 6480 to 6740, and 7590 to 7890~{\AA}) with a nominal resolving power of R=28,000. A high resolution mode (R=50,000) is available, but it suffers from more pronounced an variable optical aberrations than the lower resolution mode, as the PSF is completely dominated by the optics' aberrations. Since GALAH survey uses the low resolution mode, we use the same lower resolution for the analysis presented in this paper.

Hermes is fed by the 2dF fibre positioner, which collects light from a 2$^\circ$ diameter field of view at the AAT prime focus into 400 fibres. The light enters the fibres via a prism. They are both attached to a ``button'' which is held in place on a metal plate by a magnet. There are two plates at  the prime focus each equipped with a different set of 400 fibres. Fibres on one plate can be reconfigured while the other plate is used for observing. 

On the spectrograph end the fibres are arranged into an artificial slit (pseudo-slit) in a pattern where the mean distance between fibres is around 8 pixels when imaged on the CCD and between every ten fibres there is a larger gap of 22 pixels. 8 fibres are used for guiding and some fibres are dead, so not all of the 400 fibres are used to collect the data. Due to two sets of fibres, associated with different plates, there are also two pseudo-slits which can be moved in and out of the position inside the spectrograph. The arrangement of fibres is similar in both but not exactly the same. Fibres are fixed permanently into the two pseudo-slits, so there is only one fibre configuration observed in the CCD plane for each set of fibres. 

\begin{figure}
\includegraphics[width=\columnwidth]{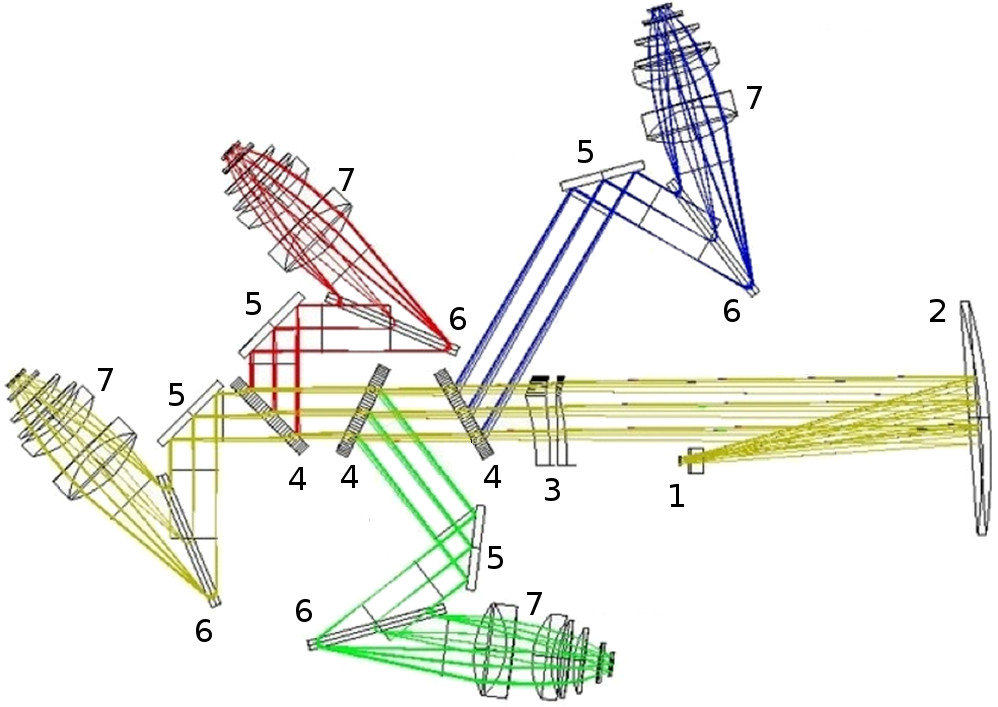}
\caption{Schematics of the Hermes spectrograph. 1: Slit assembly, 2: Collimator, 3: Corrector, 4: Beam splitters, 5: Fold mirrors, 6: Gratings, 7: Camera and CCD assemblies. Blue, green and red rays mark the respective arms and yellow rays mark the IR arm.}
\label{fig:1}
\end{figure}

The light then travels through a collimator and is split into blue, green, red, and IR arms by beam splitters. In each arm is a volume phase holographic grating in a Littrow configuration followed by a camera and a CCD (Figure \ref{fig:1}). 

Neglecting any variations in time, we have to deal with eight different settings when reducing or analysing the data: four CCDs with two different sets of fibres each.

\subsection{The photonic comb}

A key step in parameterizing the PSF of the Hermes spectrograph is using a photonic comb to measure the PSF. The photonic comb is a Fabry-Perot etalon formed in a small section of SM600 single-mode fibre with each end coated with a dielectric mirror (this section is then inserted into a longer length of single-mode fibre). The resulting transmission spectrum of the device is a set periodic peaks in frequency. Here the etalon cavity is 457~$\mu$m long and 4~$\mu$m wide. The etalon has a finesse (ratio of the peak separation to peak width) of $\sim40$. It is designed to operate in a wavelength range between 5000 and 8000~{\AA}, which unfortunately does not include Hermes' blue band. The cavity is small enough so it can be easily thermally controlled and thus stabilized to 10~$\mathrm{cm s^{-1}}$ with a Rb laser locking setup or to 10~$\mathrm{m s^{-1}}$ without the setup, like in our case \citep{betters16}. The photonic comb is used to filter an NKT SuperK COMPACT supercontinuum light source, where the resulting transmission spectrum was then observed with Hermes. The etalon can withstand sufficient high power levels to allow reasonable exposure times for our experiment (50 minutes).

A photonic comb has some major advantages over an arc lamp when used for measuring a PSF. The peaks are well separated, unlike in an arc lamp spectrum, where peaks are often blended and it is sometimes hard to establish even the magnitude of this problem. There are also big gaps in the wavelength space where no measurement could be made in an arc lamp spectrum. This causes problems for performing wavelength calibration in the GALAH survey \citep{kos17}, these problems would be even more prominent with the precise PSF measurements done in this work. The peaks in the arc lamp spectrum vary a lot in power, while in a photonic comb spectrum they only vary within $\sim20$\%. The analysis techniques can therefore be optimized for high SNR photonic comb data only. 

A major issue when using a photonic comb is the shape of the peaks. A finesse of 40 means the peaks are well separated, but the width of each peak is still resolved. Therefore a PSF cannot be measured directly, as it would be dominated by the intrinsic shape of the peaks. We have a good knowledge of the intrinsic shape of the peaks \citep{betters16} which can be used to deconvolve the intrinsic shape out of our images (detailed in Section \ref{sec:deconvolve}). Another issue are internal reflections (ghosts) that produce stray peaks in the image. They can be seen in Figure \ref{fig:deconvolution}. We were unable to determine the exact origin of these ghosts or their pattern. They are, however, not too frequent, weak, and random, so we ignore them. Their influence is eliminated by our decomposition of PSF into moments, as it is completely insensitive to random noise.

\subsection{Injecting photonic comb light into the fibres}

\begin{figure}
\includegraphics[width=\columnwidth]{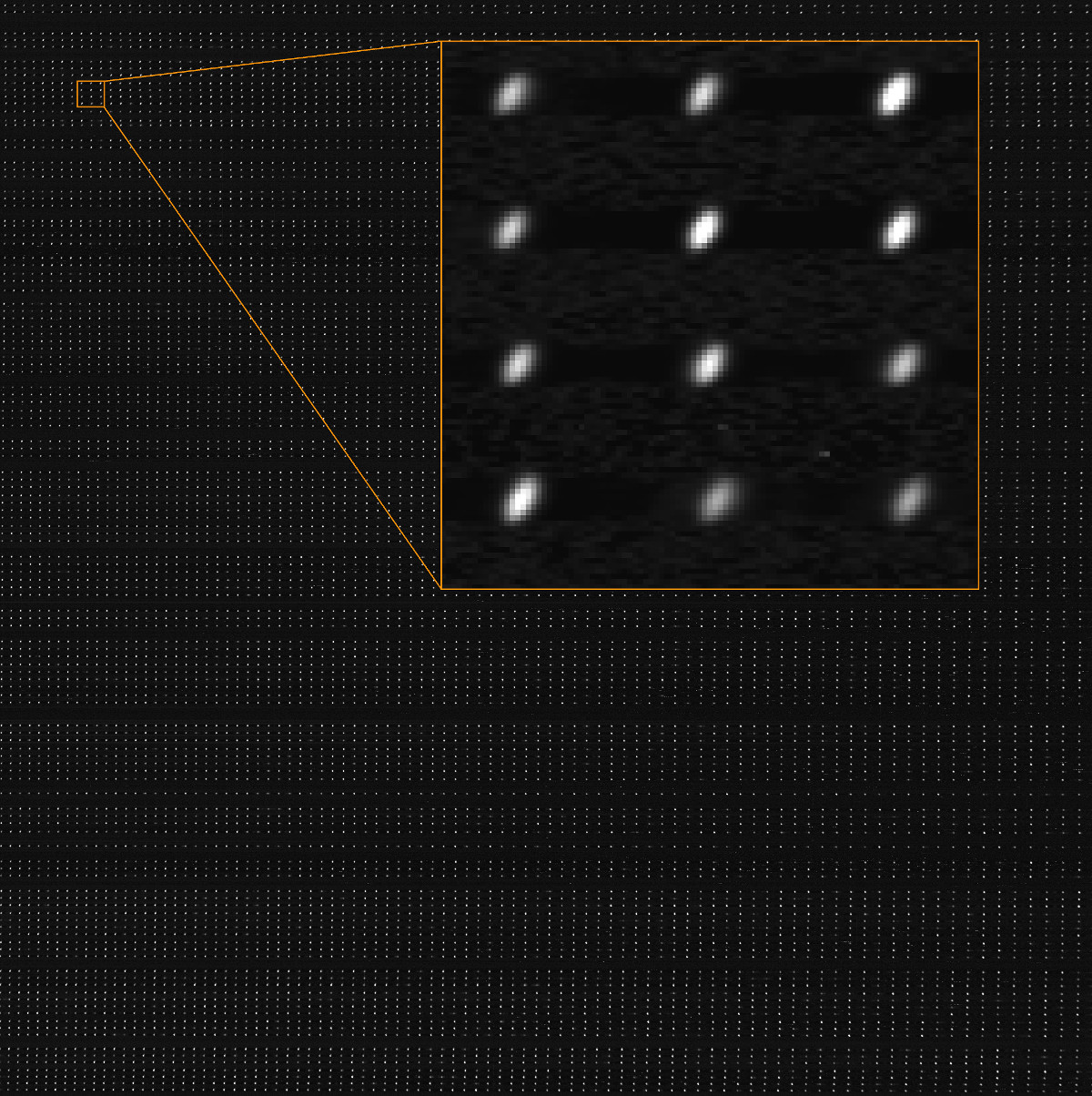}
\caption{Snapshot of a photonic comb spectrum in the green arm. Only every third fibre is illuminated. Zoomed inset shows dozen peaks in more detail.}
\label{fig:2}
\end{figure}

Ideally the light from the photonic comb would be reflected off the flat-field screen on the dome and observed through the telescope to imitate the conditions used for science exposures as well as possible \footnote{In this paper we aim to reproduce observed Solar spectra. Solar spectra are obtained by observing a twilight sky, which is, like a dome screen, evenly illuminated. Stars, on the contrary, are point sources and do not necessarily illuminate the whole fibre evenly.}. The used photonic comb is not powerful enough to do that, so we have to inject the light into the 2dF fibres directly. We injected the photonic comb light through a small 24~cm Schmidt Cassegrain telescope (SCT). The SCT was set 1.5~m in front of the 2dF plate where fibres were positioned into a 24~cm diameter field illuminated by the SCT. This assures an even illumination of the fibres, minimal light loss and a geometry that resembles real observations as well as possible. The illuminated area is large enough that half of 400 fibres can be positioned in it. The SCT telescope has a central obstruction of 6~cm, so the illuminated region is actually a ring. The light from the photonic comb is led into the SCT through a multi-mode fibre. The focal ratio of the SCT is f/10, which is matched with the focal ratio of the fibre with a lens. To avoid loosing most of the light in the central obstruction, the single-mode fibre from the photonic comb is coupled to a multi-mode fibre that leads into the SCT with a small offset. This produces an annular mode in the multi-mode fibre, so most of the light goes past the central obstruction. See Figure 1 in \citet{jbh17} for an illustration. 

To ensure that peaks are well separated in the spatial direction we only probed every third fibre at once with the exposure time of 50~minutes divided into 5 exposures. To probe all 400 fibres on both plates takes one day, including fibre positioning. One example of an imaged photonic comb spectrum is in Figure \ref{fig:2}.

\section{Analysis of the photonic comb images}

\subsection{Deconvolution with a Lorentzian profile}
\label{sec:deconvolve}

The intrinsic shape of the peaks produced by the photonic comb is the Airy function:
\begin{equation}
T(\lambda)=\frac{1}{1+F \sin^2\left( \frac{2 \pi n l}{\lambda} \right)},
\end{equation}
where $n$ is the refractive index of the cavity with size $l$, and $F$ is the finesse. Although in practice, the peaks have different strengths. With the finesse of around 40, the peaks are well separated, but still resolved with detectable flux in-between the peaks. The observed peaks are therefore not a good representation of the instrument's PSF. We use the Richardson-Lucy \citep{richardson72,lucy74} algorithm to deconvolve the observed image with an appropriate kernel. Since the Airy function is periodic, we approximate the intrinsic shape of a peak with a Lorentzian kernel, which is a very good match for an individual peak in the Airy function:
\begin{equation}
L(\lambda)=\frac{1}{2\pi}\frac{\Gamma}{\left(\lambda-\lambda_0\right)^2 + \left(\frac{1}{2}\Gamma\right)^2},
\end{equation}
where $\lambda_0$ is the position of the peak and $\Gamma$ is the width parameter. Because the orientation of the spectra in the image is very well aligned with the pixels grid (spectra are tilted or twist at most for one pixel in the vertical direction for every thousand pixels in the horizontal direction), we can perform the deconvolution on the raw image in the wavelength direction only, without tracing the spectra first. 

The Richardson-Lucy algorithm works by iteratively finding the signal that when convolved with the kernel best represents the observed signal. It assumes Poissonian statistics for the signal and is not sensitive to noise. Our problem is simple with an elementary one-dimensional kernel, so the solution can be found quickly. The dispersion, however, varies through the spectra, so the width of the Lorentzian kernel (in pixels) depends on the position in the image. The variation is known, is small, smooth, and the same in all fibres (the dispersion changes between the fibres for no more than 0.1\%). The problem can be solved by simply splitting the images into vertical slices and using a kernel with the appropriate width for each slice. 

The Lorentzian kernel has only one free parameter -- the width $\Gamma$ -- which we have to find. It is provided by the manufacturer of the photonic comb, and was further checked and fine tuned by us by comparing deconvolved images where different widths were used. A good solution must converge quickly, produce no artifacts, like negative fluxes around peaks and remove the obvious Lorentzian wings. The deconvolved PSF can also be compared to the images of the arc lines in selected well-behaved regions, as the PSFs should look the same. Intrinsic widths of arc lines are several orders of magnitude smaller than the width of Hermes' PSF, so they do not have to be deconvolved. The problem is that some lines are blended, so a clean region where a fair comparison can be made must be chosen.

\begin{figure*}
\includegraphics[width=\textwidth]{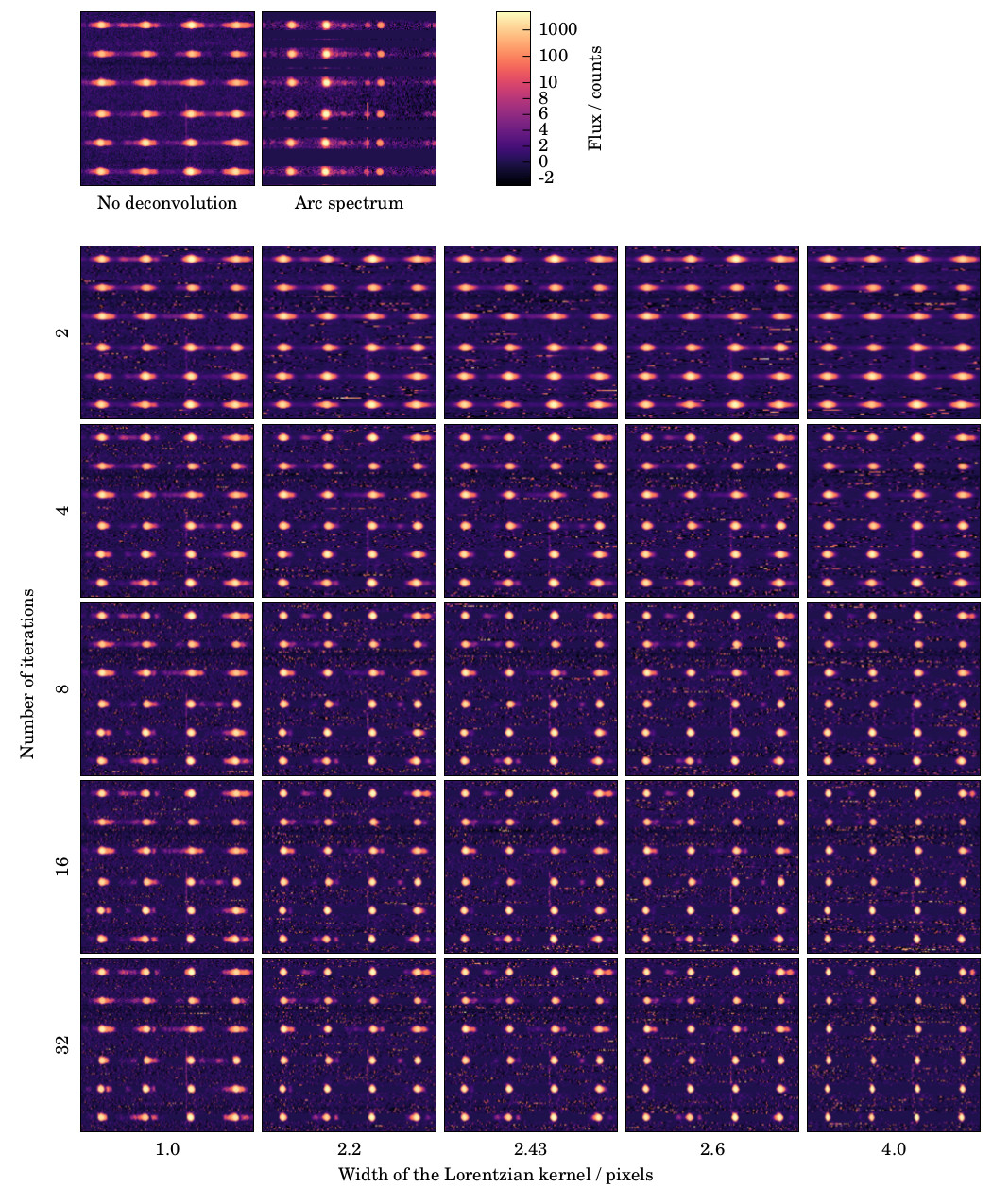}
\caption{A small region of the CCD plane is shown, deconvolved with Lorentzian kernels of different widths (left to right) and with a different number of iterations used (top to bottom). For reference, in the top, the same region before the deconvolution is plotted, as well as an arc lamp spectrum. Arc spectrum is produced with all of the fibres illuminated, so we took the liberty to mask out spectral traces that are missing in other panels. Each panel is $170\times 170$ pixels in size and shows the part of the green arm CCD plane with least amount of distortions. Note the nonlinear flux mapping that enhances how the low-count pixels are displayed.}
\label{fig:deconvolution}
\end{figure*}

Figure \ref{fig:deconvolution} shows a small part of the green arm CCD image deconvolved with different width parameters and different number of iterations. The deconvolution is not very sensitive to the width of the kernel and the solution is stable after $\sim 15$ iterations with variations of $\ll1\%$. The width must be decreased considerably before the deconvolution fails to remove the Lorentzian wings, or increased significantly  before the solution does not converge any more. An artifact is also included in the displayed region, a trace of a cosmic ray, demonstrating that it does not have any impact on the deconvolution process. The final adopted kernel width is $\Gamma=2.43$~pixels for the centre of the green arm.

\subsection{Image moments}

First, we would like to analyse the measured PSFs and learn what kind of optical aberrations we can see and if there are any patterns appearing over the field. To analyse the PSFs the image moments are used, as they have intuitive meanings.

For a 2-dimensioanl image $f(x,y)$ of size $N\times N$ the image moments of order $p,q$ are defined as:
\begin{equation}
T_{pq}=\sum_{x=0}^{N-1}\sum_{y=0}^{N-1}x^py^qf(x,y).
\end{equation}
In this work we prefer to use central image moments, which are invariant to translation, defined as:
\begin{equation}
\mu_{pq}=\sum_{x=0}^{N-1}\sum_{y=0}^{N-1}(x-\bar{x})^p(y-\bar{y})^qf(x,y),
\end{equation}
where $(\bar{x}, \bar{y})$ is the centre of the centroid. This way the position of the analysed PSF peak does not have to be known with a subpixel accuracy.

Image moments conveniently have intuitive meanings:
\begin{itemize}
\item $\mu_{00}$ is the power in the image,
\item $\mu_{22}/\mu_{00}$ is the overall variance of the image,
\item $\mu_{33}/\mu_{00}$ is the overall skewness of the image,
\item $\mu_{44}/\mu_{00}$ is the overall kurtosis of the image,
\item $\frac{1}{2}\arctan\left( \frac{2\mu_{11}/\mu_{00}}{\mu_{20}/\mu_{00}-\mu_{02}/\mu_{00}} \right)$ is the rotation of the dominant image axis.
\end{itemize}
We will also use two Hu \citep{hu62} moments:
\begin{equation}
I_1=\eta_{20}+\eta_{02},\\
I_2=(\eta_{20}-\eta_{02})^2+4\eta_{11}^2,
\end{equation}
where
\begin{equation}
\eta_{pq}=\frac{\mu_{pq}}{\mu_{00}^{1+(p+q)/2 }},
\end{equation}
so the $I_1 / I_2$ is associated with the roundness of the image.

Unfortunately there is no easy way to reconstruct the original image from the image moments, as they are not orthogonal. A method called ``moment matching'' \citep{teague80}, which is only feasible for a reconstruction from a small number of moments is too slow to use in a real-life application. It requires one to solve an increasingly larger system of linear equations for every used moment. Reconstruction via a Fourier transform is simpler, but it requires one to calculate more moments than the previous method. The formula reads:
\begin{equation}
F(u,v)=\sum_{m=0}^{N_\mathrm{max}}\frac{(-2i\pi)^m}{m!} \sum_{k=0}^m {{m}\choose{k}} \left( \frac{u}{N} \right)^{m-k} \left( \frac{v}{N} \right)^k \mu_{m-k,k},
\end{equation}
where $N_\mathrm{max}$ is the highest moment we intend to use. \\${{m}\choose{k}}=\frac{m!}{k! (m-k)!}$ denotes a binomial coefficient. Reconstructed $f(x,y)$ is then the inverse Fourier transform of $F(u,v)$. We deem the reconstruction from image moments computationally too slow to serve our needs, especially when many other moments, like discrete Chebyshev moments (Equation \ref{eq:division}) are better suited.

\subsection{Decomposed optical aberrations}

\begin{figure*}
\includegraphics[width=0.95\textwidth]{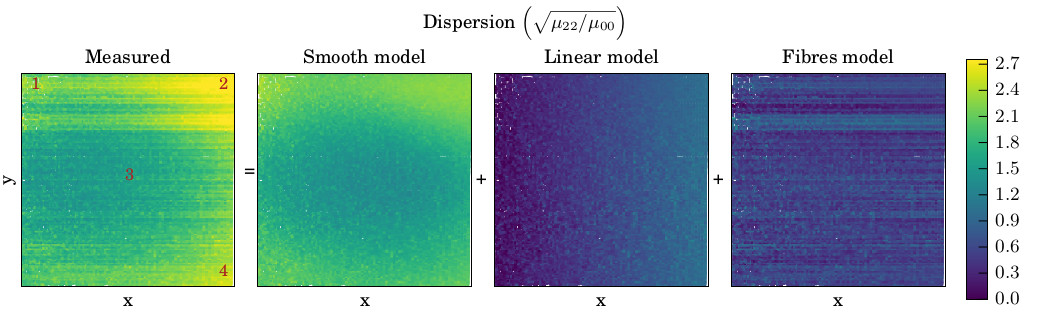}\\[0.52cm]
\includegraphics[width=0.95\textwidth]{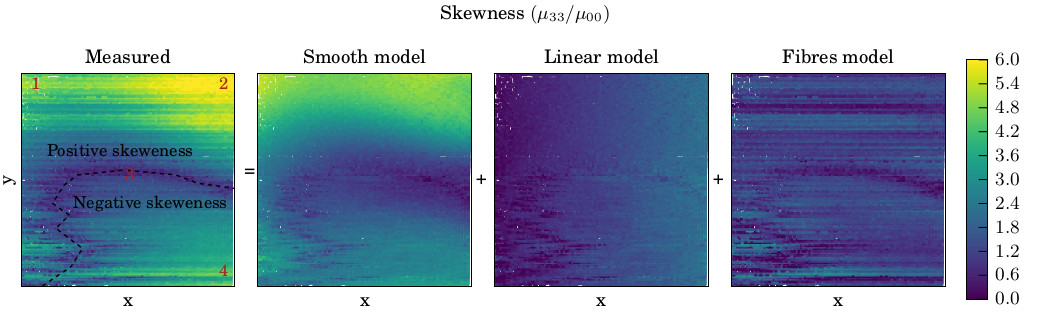}\\[0.52cm]
\includegraphics[width=0.95\textwidth]{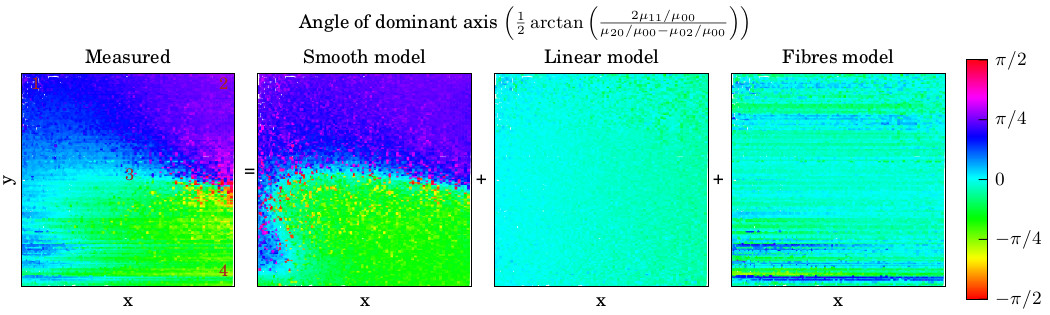}\\[0.52cm]
\includegraphics[width=0.95\textwidth]{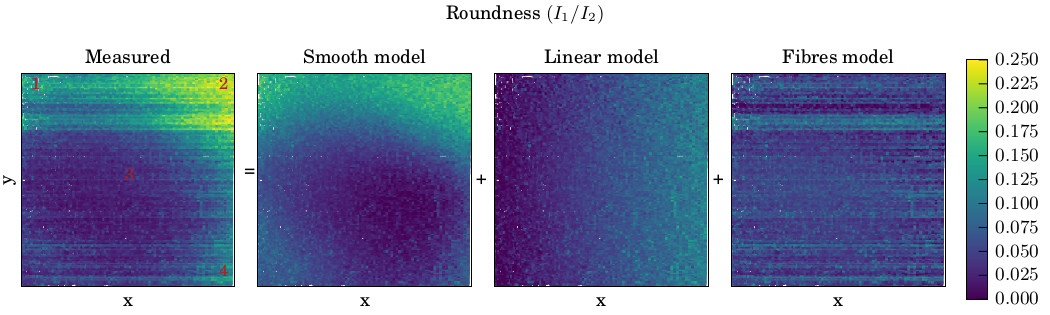}
\caption{Left-most panels show the measured properties derived from moments in the whole image. Each measured PSF is displayed with a colour-coded point in each panel. Left to right follow decomposed components. Instead of plotting the fitted component, the moments are plotted with the other two components were removed. This way we can also show how noisy each component is. Top to bottom follow four different properties calculated from the moments. See also Figure \ref{fig:decomposed_green_cutout} to visualize the PSFs in the regions marked by numbers 1-4 in the above plots.}
\label{fig:decomposed_green}
\end{figure*}

Different optical aberrations can originate in different parts of the optics. It is possible to distinguish the aberrations if their influence shows a distinct variation over the CCD or if they affect only certain moments. Figure \ref{fig:decomposed_green} displays the decomposed image moments in the green arm. Looking at the panels in the first column, one can distinguish a smooth component and a component that changes sporadically from fibre to fibre (along the y axis). The latter component comes from fibres that can have different apertures, so they produce a beam with a different PSF. Fibres have a major influence on the shape of the PSF, but negligible influence on the orientation, for example. The smooth component can be decomposed further into a linear component, and a spherical component. The linear component shows changes along the wavelength axis only. This is associated with the aberrations that the gratings can produce. The remaining part of the smooth component is nearly spherically symmetric. This component is associated with the optics, as Hermes optics consists of mostly round elements. Note that there is a certain level of degeneracy between all three components. The spherically symmetric smooth component also includes some offset, while the other two components are represented as the correction on top of the former component. The offset could be instead distributed between any of the three components, so the exact amount of aberration contributed by each component is not known.

\begin{figure}
\begin{minipage}{0.49\columnwidth}
\centering
\includegraphics[width=\textwidth]{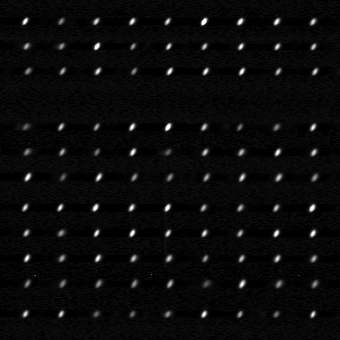}\\
(1)
\end{minipage}
\begin{minipage}{0.49\columnwidth}
\centering
\includegraphics[width=\textwidth]{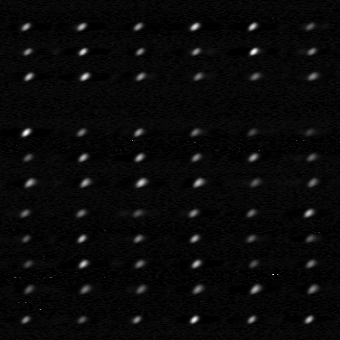}\\
(2)
\end{minipage}\\[1mm]
\begin{minipage}{0.49\columnwidth}
\centering
\includegraphics[width=\textwidth]{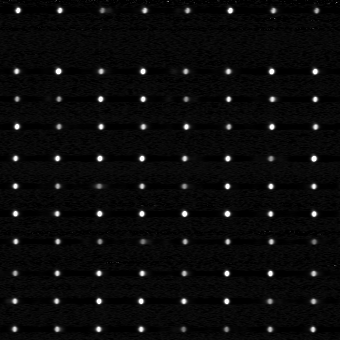}\\
(3)
\end{minipage}
\begin{minipage}{0.49\columnwidth}
\centering
\includegraphics[width=\textwidth]{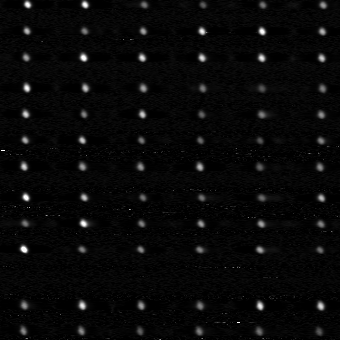}\\
(4)
\end{minipage}
\caption{Small regions of the deconvolved image showing photonic comb peaks in the green arm when every third fibre is illuminated. Cut-outs 1-4 correspond to positions marked in Figure 
\ref{fig:decomposed_green}.}
\label{fig:decomposed_green_cutout}
\end{figure}

\section{Image reconstruction}
\label{sec:section4}

\subsection{Chebyshev moments}
\label{sec:moments}

For the purpose of image reconstruction we use moments based on orthogonal functions. There are many appropriate orthogonal moments. Here we use discrete Chebyshev moments for their good performance and convenience. A base function for the moments is a discrete classical Chebyshev polynomial of order $n$ defined as:
\begin{equation}
t_n(x)=n!\sum_{k=0}^n (-1)^{n-k} {{N-1-k}\choose{n-k}} {{n+k}\choose{n}} {x\choose k},
\label{eq:def}
\end{equation}
where $N$ is the size of a discrete range (one dimension of an image in pixels). This constrains the number of possible discrete polynomials to $n=0,1,2,\ldots, N-1$. The following condition must also be satisfied:
\begin{equation}
\sum_{x=0}^{N-1} t_p(x)t_q(x)=\rho(p,N)\delta_{pq},\quad 0\leq p,q\leq N-1
\end{equation}
where
\begin{equation}
\rho(p,N)=\sum_{x=0}^{N-1}\left [ t_p(x)\right ]^2 = \frac{N(N^2-1)(N^2-2^2)\cdots(N^2-p^2)}{2p+1}
\end{equation}

Chebyshev moment of order $p,q$ of an image with size $N \times N$ and intensity $f(x,y)$ is defined as:
\begin{equation}
T_{pq}=\frac{1}{\rho(p,N) \rho(q,N)}\sum_{x=0}^{N-1} \sum_{y=0}^{N-1} t_p(x) t_q(y) f(x,y).
\label{eq:division}
\end{equation}
Like for the base functions, only moments where $p,q<N$ exist.

The inverse transformation is given as:
\begin{equation}
f(x,y)=\sum_{p=0}^{N_\mathrm{max}}\sum_{q=0}^{N_\mathrm{max}} T_{pq} t_p(x) t_q(y),
\end{equation}
where $N_\mathrm{max}$ is the highest moment we intend to use. Moments up to and including $N-1$ can be used, following the definition in Equation \ref{eq:def}.
Discrete Chebyshev moments are defined for images of any size, not just $N \times N$. The above simplification is used, because we only deal with square images in this work.

Note that discrete Chebyshev moments are exact, as only whole numbers enter the formulas above. Following a division in Equation \ref{eq:division} we have to deal with rational numbers, which in practice limit the precision, but since only a few operations are done before the image is reconstructed, the loss of precision is negligible. Because only a finite number of moments exist, the image can be decomposed into moments and reconstructed without any loss of information, as long as all existing moments are calculated and used. This is a huge advantage over Zernike or Legendre moments, often used in similar analysis \citep{mukundan98}. We do not want to approximate our reconstructed image, so we use all of the existing moments. Decomposition into moments is more of an interpolation algorithm in our case.

Chebyshev moments also require no coordinate transformation and work in the original pixel space of the image. Most other base functions require a normalization of the image into a uniform range, usually $[-1:1]$. Working in the original pixel space is convinient, although not a big advantage, as we still have to reinterpolate the PSFs. Because the discrete Chebyshev moments are not translation invariant, we have to calculate the centre of each PSF and reinterpolate them, so the centre is aways in the middle of the central pixel. Finding the centre is not trivial. We take great care that the same algorithm and parameters are used as for finding the centres of spectral traces (or tramlines). See Section \ref{sec:tramlines} for details on fitting the tramlines.

Since we are dealing with relatively small image sizes (cutouts of 15$\times$15 pixels large), the base functions $t_n(x)$ can be tabulated for all possible combinations of $n$, $x$, and $N$ and queried quickly, which significantly reduces the computation time.

It has also been shown that Chebyshev moments are less sensitive to noise \citep{mukundan00} than Legendre and Zernike moments, which is an important feature when dealing with astronomical data.

\subsection{Interpolation of moments}
\label{sec:interp}

To reconstruct an image the PSF in each and every point on the CCD plane must be known. Therefore the measured PSFs between the peaks that the photonic comb produces must be interpolated. Individual measured PSFs are also affected by noise and cosmetic artifacts, so individual measured PSFs cannot be taken for granted. Both problems are solved by parameterizing the measured PSFs with discrete Chebyshev moments and fitting a model to each moment. 

A model for the power of each moment in the CCD plane consists of a smooth component and a fibre component. For a smooth component we use a two-dimensional, 5$^\mathrm{th}$ order Chebyshev polynomial of the first kind in this work. The exact function used is not essential, and the order can be determined by inspecting the modeled moments. There is no local variation in any moment that could not be described by a 5$^\mathrm{th}$ degree polynomial. This is not true for fibre-to-fibre variations. These variations are modeled as a correction to the smooth fit. The fibre correction is a 3$^\mathrm{rd}$ degree polynomial tracing the variation in each moment along the wavelength axis. Each fibre is treated separately and independently, so the correction for each fibre can be different. A sum of the smooth model and all fibre models gives us the interpolated model for the power of each moment anywhere in the CCD plane. Even more, if fibres are shifted in the CCD plane, we can still calculate the correct model by shifting the fibre component of the model against the smooth model. In practice the shifts are only of a few pixels, so the effect is almost negligible.

Usually orthogonal moments are used when one wants to know only a general shape of the image given by a few lowest order moments. Only low-order moments could be used to reconstruct the PSF for every spectrum we reproduce, but then some information about the PSF would be lost. Since we parametrized the PSF with moments and produced a smooth model for each moment, a smooth model for the PSF itself can be produced. This only has to be done once for each image, so we can afford to use all the moments and produce a PSF without any loss of information. This is also sensible time-wise, as during the reconstruction we do not have to calculate the PSF from the moments at every iteration for every pixel. A PSF at any point on the CCD plane is now represented by an image $PSF(x',y';f)$ of, in our case, $15\times15$ pixels, where value in each pixel is given by a smooth model:
\begin{equation}
PSF(x',y';f)=A_{p,q}(x,y)+B_{p,q}(x,f), \qquad p,q=[0..14].
\end{equation}
Functions $A_{m,n}$ for the smooth and $B_{m,n}$ for the fibre component of the model are calculated from the moments.
Fibre part of the model depends only on coordinate $x$, if tramlines are aligned horizontally, and, of course, on fibre $f$. $PSF(x',y';f)$ is normalized, so the total flux in the PSF is 1 everywhere on the CCD plane. This is achieved by setting moment $0,0$ to 1. 

The power of all 225 discrete Chebyshev moments in the entire CCD plane is shown in Appendix \ref{sec:app_all}.

\subsection{Preparations for the image reconstruction}

Before we attempt to reconstruct an image, a few steps of image reduction must be addressed that are not included in the reconstruction process or are treated the same as in a traditional reduction.

\subsubsection{Bias, dark, flat field}

Bias, dark and flat fields can be included in the reconstruction or treated the same as in the traditional extraction. The advantage of including them into the reconstruction is that they can be modeled, for example in a probabilistic way \citep[e.g.][]{burger11, harpsoe12}. The aim of this paper is to demonstrate analysis of spectra through image reconstruction, so we used a traditional reduction in this case. Bias was removed and the damaged columns that can be identified from a flat field were corrected. 

\subsubsection{Scattered light and stray light}

The PSF, which is represented by $15 \times 15$ pixels does not contain all of the light from a monochromatic point source. Some of the light is scattered and illuminates the CCD far away (much more than 15 pixels) from the centre of the PSF. Gaps between sleetlets are wide enough that the scattered light can be measured and approximated by a low order polynomial. It was subtracted from our science images in the same way as in a traditional reduction \citep{kos17}. 

Any additional scattered light can be picked out in the residual image -- a difference between the observed and reconstructed image as shown in Section \ref{sec:recovery}

\subsubsection{Littrow ghost}

The gratings in the Hermes spectrograph are in a Littrow configuration which produces a ghost -- a non-dispersed reflection of the slit close to the middle of the image. Because the pseudoslit has a non-trivial fibre arrangement it is hard to model the shape of the ghost. The ghost is weak, several orders of magnitude weaker than the spectrum itself and only corrupts a few pixels in each spectrum. It is not a major aberration, so we ignore it for the purpose of this study.

\subsubsection{Tramlines}
\label{sec:tramlines}

Before we attempt to reconstruct the image, we have to know where on the image spectral traces, or tramlines, lie. Tramlines do not necessary lie along the same curves on every image. The pseudo-slit is usually moved between different exposures, and if its position changes, all the tramlines are shifted. This effect is usually very subtle, but not negligible. Our tramlines have a width of FWHM=4 pixels. If we reproduce the spectrum along a tramline that is shifted by just $3/100$ of a pixel,  this can lead to  differences   of up to 1\% between the original and reproduced image.  It is therefore best to measure the tramlines on the image we are trying to reproduce.

\begin{figure}
\includegraphics[width=0.95\columnwidth]{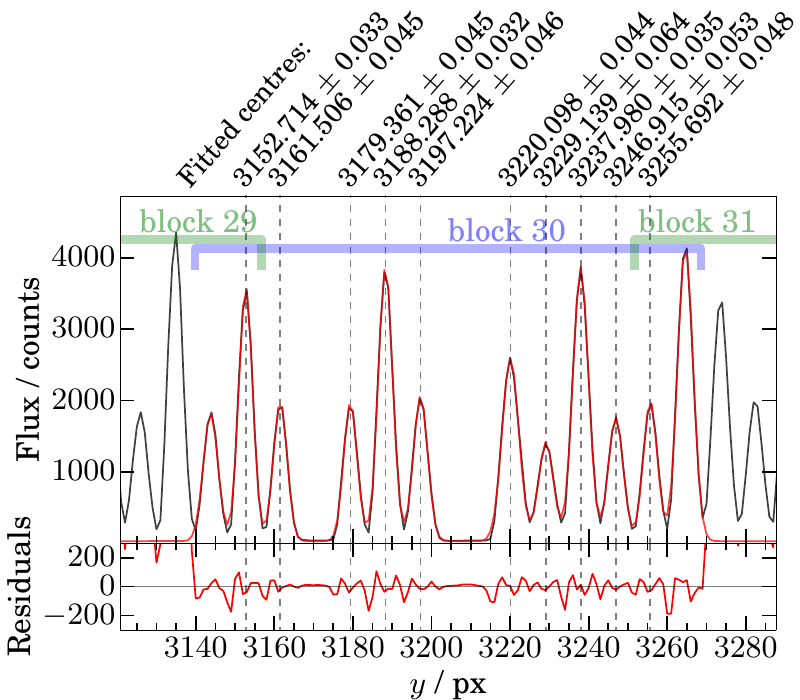}
\caption{Centres fitted to cross-sections of 10 consecutive tramlines. Top: Cross-section is plotted in black and the best fit is plotted in red. Note that 12 centres (in block marked with blue) are fitted, but the first and the last one are rejected. They will be calculated together with centres marked as green blocks. Measured centres are printed on top. Bottom: residuals between the fit and the cross-section.}
\label{fig:cross_section}
\end{figure}

Measuring the tramline positions with $1/100$ pixel accuracy is not trivial. The centre of the tramline must be measured at many positions along the $x$ axis, then the tramline curve should be fitted through measured centres. We make a cross-section of the whole image at 10 pixels intervals and average the flux in such 10 pixels wide blocks. Higher signal will help us measure the centre of each tramline in the cross-section. In such cross-section we measure centres of all tramlines (the actual number depends on how many fibres are illuminated, usually there are around 380 usable fibres). Centres are found by fitting  a Gaussian to each peak and adopting the measured mean as a tramline centre at a given coordinate $x$. Tramlines are packed close enough that neighbouring tramlines interfere with the fit for a single tramline. Therefore we never fit only one tramline, but several (10, for example) at the same time, as illustrated on Figure \ref{fig:cross_section}. We create a model of 12 Gaussians (36 free parameters), representing 12 consecutive tramlines and fit the model to the given cross-section. This way we get 12 centres for 12 tramlines. Measurements for the first and last tramline are discarded, as they are influenced by the neighbouring tramlines that were not included in the model. We are left with reliable measurements for 10 consecutive tramlines. The process is then repeated for the next 10 tramlines. A block of 10 tramlines was chosen because this is most efficient. If more are fitted at the same time, the fit would take too long to converge. Since two measurements are discarded every time, they have to be repeated, so to minimize the number of repetitions as many tramlines as can be reliably fitted are processed at the same time. 

With the described approach, the centre measuring accuracy is limited by the SNR and the sampling. We increase SNR by averaging 10 pixels wide blocks. Unfortunately nothing can be done about poor sampling along the vertical direction. A Gaussian must be fitted to a cross-section where each peak is represented by $\sim8$ points.

Figure \ref{fig:cross_section} shows one example of fitting the centres. In the shown case the uncertainty of each centre is of the order of $0.04$ pixel. The actual uncertainty of the tramline is lower than that, because the tramline is a low order polynomial fitted to around 400 such measurements. 

If the fitted tramline is not accurate enough, the mismatch will be obvious in the residuals from the reconstructed image. This mismatch is measured and used to correct the tramlines fits.

\subsubsection{Fibre throughputs and blaze function}

A reconstructed image is produced from a normalized template. The observed image, however, carries all the information of fibre throughputs, blaze function and spectrograph response. These could be measured from the photonic comb, if it were properly calibrated and stable. The power of individual peaks in our photonic comb is neither stable or calibrated, hence we omit moments (0,0) in the analysis in Section \ref{sec:interp}. The response of the optics depending on the wavelength must thus be measured from science images themselves. The continuum can be measured from our observed spectra by making a simple extraction. A measured continuum is then used to denormalize the synthetic spectrum before it is convolved with the PSF. A few iterations have to be made, so the continuums measured on the observed image and reproduced image are the same, assuming the best matching template was used. In our case the continuums are represented by a 15$^\mathrm{th}$ degree polynomial. We call the obtained function a denormalization function, as it is used to rescale a previously normalized template.

\subsubsection{Telluric spectral features}

Hermes bands are moderately affected by telluric absorption and emission lines, avoiding the strongest few but containing a large number of weaker lines. These must be added to the template spectrum, which is not trivial. The strength of the telluric lines varies with weather and zenith angle. The strength also varies within the 2 degree wide field in which the fibres can be positioned \citep{kos17}. O2 and H$_2$O are responsible for all absorption lines, including the lines in the green arm, used for demonstration here.

Here we use HITRAN database \citep{gordon17} and HAPI python module \citep{kochanov16} to calculate a transmittance spectrum of the atmosphere. In this study we only analyze a single frame of twilight spectra, so the telluric absorption spectrum has to be calculated only once. In a general application one would have to find the correct telluric spectrum in a similar way as finding the correct template spectrum. Although the telluric spectrum has only a few parameters (the amount of each absorber in the line-of-sight). The impact of absorption telluric bands is demonstrated in Section \ref{sec:template_match}.

It is harder to predict the atmospheric radiation model, as it varies more and has more sources, including  possible light pollution. Traditionally the sky emission lines are removed with the sky subtraction. With wide field spectrographs, like Hermes, this is not a trivial step. It is, however even harder to do it within our schema, as the sky spectrum would have to be known before the actual spectra are reproduced. The sky spectrum (scattered moonlight and atmospheric radiation) can be modeled, but that introduces a lot of additional free parameters. An alternative is an iteration where the best current sky spectrum is extracted and is improved once better templates for other spectra are found. 

In our case there are no dedicated sky fibres, as all the fibres observed the twilight sky. The required exposure time was low enough and the spectra were taken while the Sun was fairly quiet, so even the strongest sky emission lines are undetectable. Therefore we do not perform any sky subtraction on the data in this paper.

\section{Reconstruction of a series of solar spectra}

A twilight flat is an exposure of the sky during the morning or evening twilight. Usually it is taken as a calibration image, as the same spectrum of scattered solar light should be produced from every fibre. The sky must be reasonably bright so a short exposure time can be used in order to avoid collecting the light from the stars that might align with the fibres. The result is a series of high SNR spectra of the same source. Despite this, the traditionally reduced and extracted spectra will not be the same, as every fibre experiences different optical aberrations. 
The goal of this exercise is to show that we can take this correctly into account in our forward modeling approach. We then expect a single solar template spectrum to be equally good solution to all observed spectra.

\subsection{Recovery of misfitted parameters from the residual image}
\label{sec:recovery}

Scattered light, misfitted continuum and tramline offset can all be measured from the residual image in a few iterations and corrected retroactively. Assuming that the cross-section (along the $y$ axis) of each spectrum trace is a Gaussian (the actual cross-section is indeed very close to a Gaussian), the residual image (the difference between reproduced and original images) has a cross-section that can be described by a difference of two Gaussians:
\begin{equation}
r(y)=\underbrace{a\exp\left( \frac{-(y-y_s)^2}{2\sigma^2} \right)}_{\substack{\text{Reproduced} \\ \text{cross-section}}}-\underbrace{C b\left( \exp\left( \frac{-y^2}{2\sigma^2} \right)+f \right)\left( \frac{b}{b+f} \right)}_{\text{Original cross-section}},
\label{eq:off}
\end{equation}
where $a$ and $b$ are the reproduced and original fluxes at the centre of the tramline. $y_s$ is the shift between the true and measured tramline, $C$ is the level of misfitted continuum, and $f$ is the amount of scattered light, assuming the scattered light is uniform over the whole cross-section. $\sigma$ is the standard deviation of the Gaussian calculated by approximating the PSF with a Gaussian.

In the perfect reproduction, the residuals would be Poissonian noise only. If the fitted tramline has a small offset from the true value, so the spectrum is reproduced along a wrong tramline, the residuals are positive on one side of the tramline and negative on the other. If there is some scattered light in the original image, there are negative residuals on both sides of the tramline. The centre of the tramline should have average residuals of zero, unless the fitted continuum is off. Equation \ref{eq:off} describes a combination of all of the  above three possibilities. If we fit it to the cross-section, we get the scattered light (that is added to the PSF), shift of the tramline (for which the previously used tramline is corrected), and the continuum correction (which is used to fix the denormalization function). 

\begin{figure}
\includegraphics[width=0.95\columnwidth]{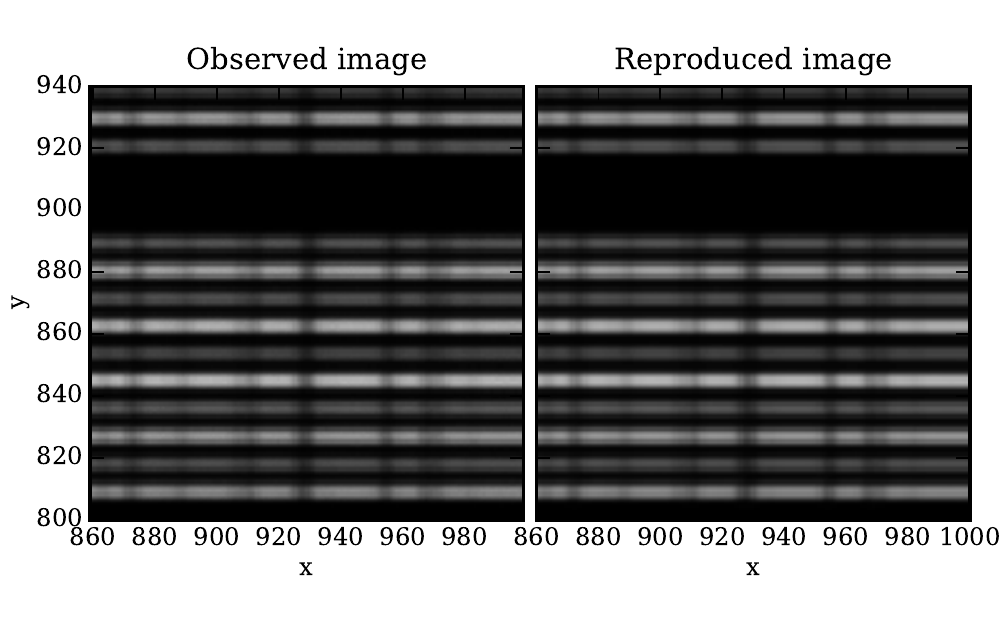}
\caption{A small part of the observed (left) and the reproduced image (right). The difference is indistinguishable by eye. See Figure \ref{fig:residuals_detail} for a more detailed representation of the differences.}
\label{fig:reproduced_dif}
\end{figure}

\begin{figure}
\includegraphics[width=0.95\columnwidth]{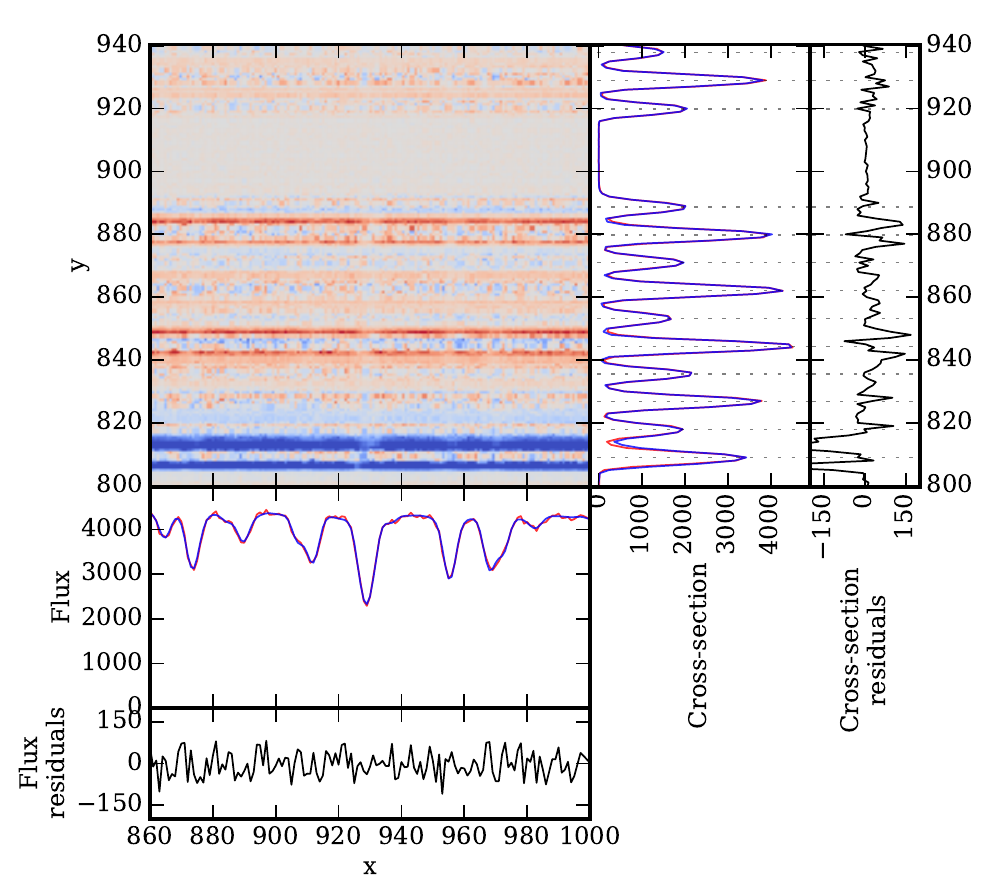}\\
\includegraphics[width=0.95\columnwidth]{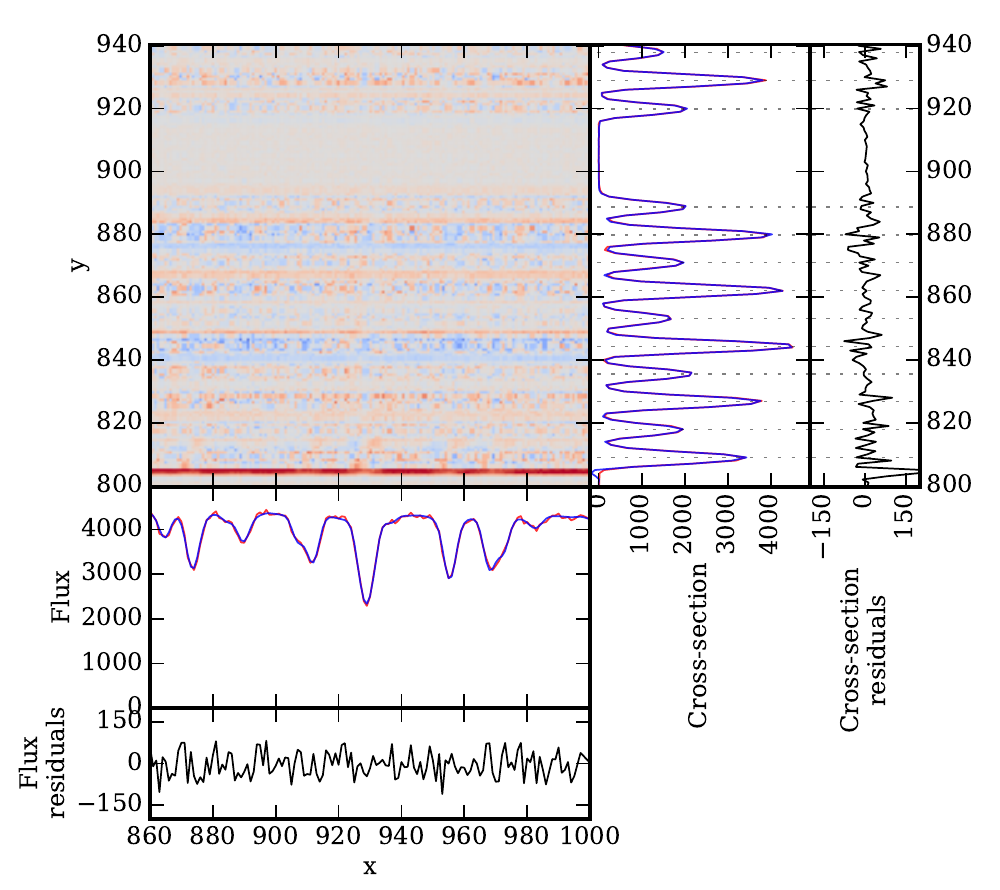}\\
\includegraphics[width=0.95\columnwidth]{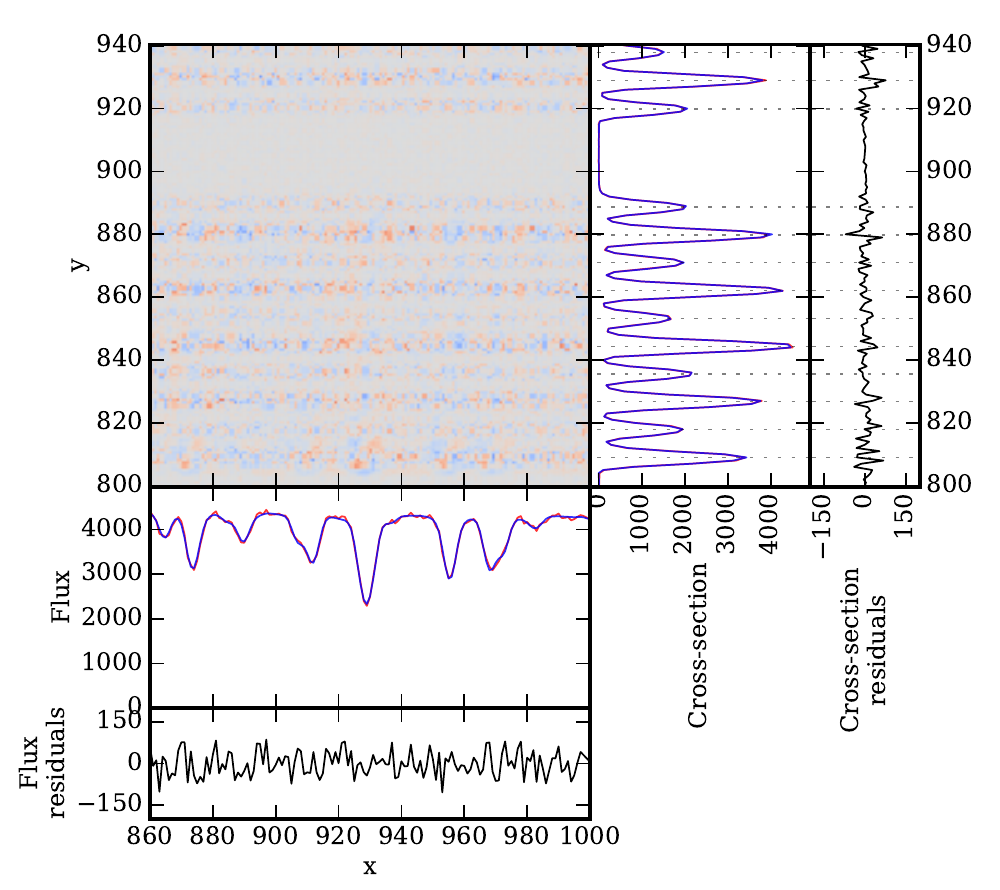}
\caption{Residuals after subtracting the reproduced image (Figure \ref{fig:reproduced_dif}, right) from the observed image (Figure \ref{fig:reproduced_dif}, left). Adjacent panels show the cross-sections of the observed (red), reproduced (blue), and the residual image (black) along the row y=862 and column x=940. Top: Residuals before tramlines, scattered light and continuum are refitted as described in Section \ref{sec:recovery}. Middle: Some horizontal features remain after the tramlines, scattered light and continuum are refitted. These are a consequence of a finite PSF size. Bottom: Horizontal features can be filtered out to produce a clean image of residuals.}
\label{fig:residuals_detail}
\end{figure}

\begin{figure}
\includegraphics[width=0.95\columnwidth]{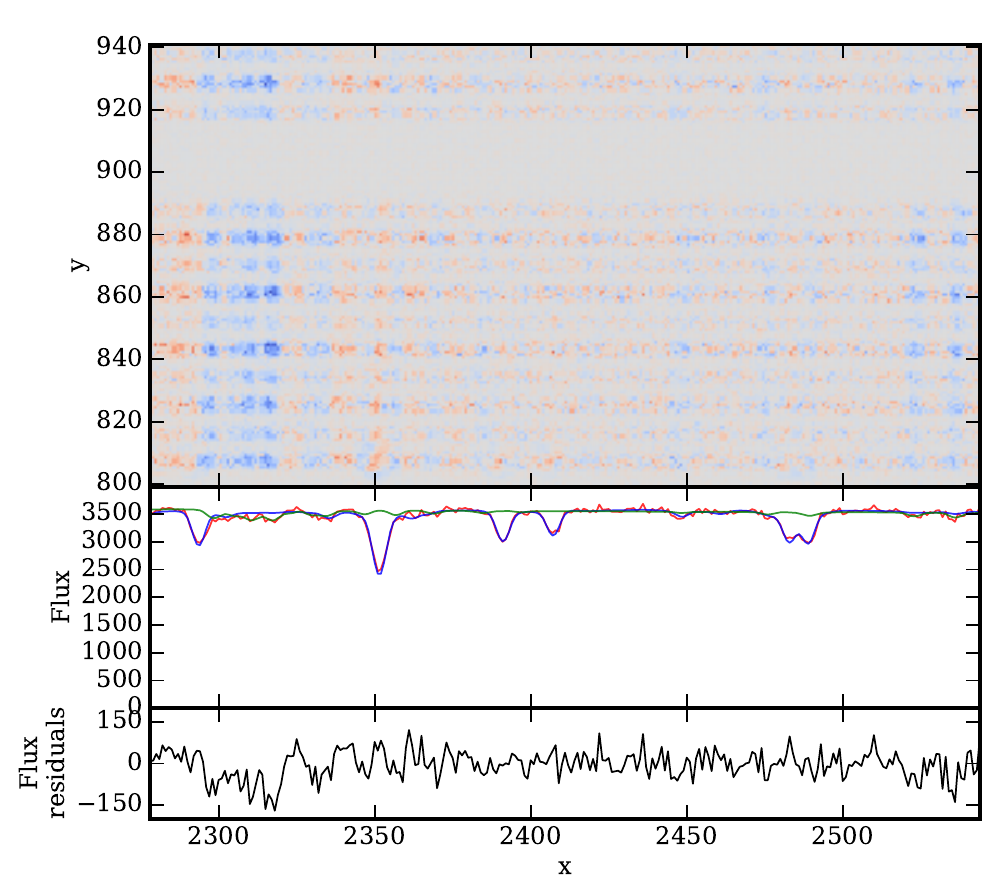}\\
\includegraphics[width=0.95\columnwidth]{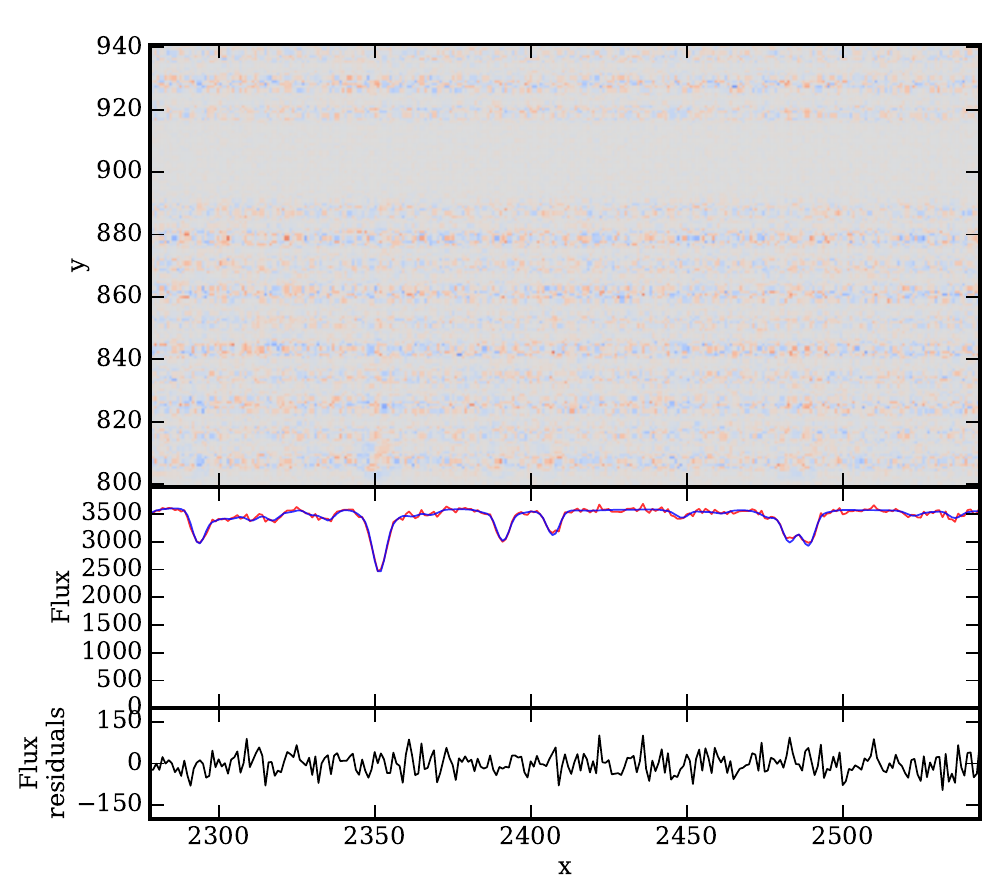}
\caption{Effect of the O$_2$ telluric bands. Residual image is shown in the same manner as in Figure \ref{fig:residuals_detail}, but without the cross-section on the right. Top: Telluric bands were not taken into the account, so they dominate in residual image. Green spectrum shows the missing telluric component. Bottom: Residuals with telluric bands taken into the account. Here the noise dominates the residuals.}
\label{fig:residuals_tel}
\end{figure}

\subsection{Template matching}
\label{sec:template_match}

With optical aberrations properly taken into the account, we want to test the performance by finding best matching Solar templates for each spectrum in our image. We expect the same template to be equally good match to the observed spectra regardless of the position on the CCD plane. A spread of well matching templates for an individual spectrum should also be reduced. 

We use a template produced from Fourier transform spectrograph scans with modeled and calculated telluric features removed \citep{kurucz06}. We did add the telluric absorptions back, but with the correct intensity. It must be noted that the twilight flats were taken at a small zenith angle, so the telluric bands are equally strong everywhere in the field. The template is normalized and comes in a resolution of 0.005~{\AA}. For a first test the template was scaled in 0.2\% steps between 85\% and 115\% to produce a grid of templates with a varying strength of spectral lines:
\begin{equation}
f_{s}=e^{s\log\left( f \right)},
\end{equation}
where $f_s$ is a scaled flux $f$, and $s$ is the scaling factor (between 0.85 and 1.15). The image was reproduced with every scaled template in our grid. 

\begin{figure*}
\includegraphics[width=0.95\textwidth]{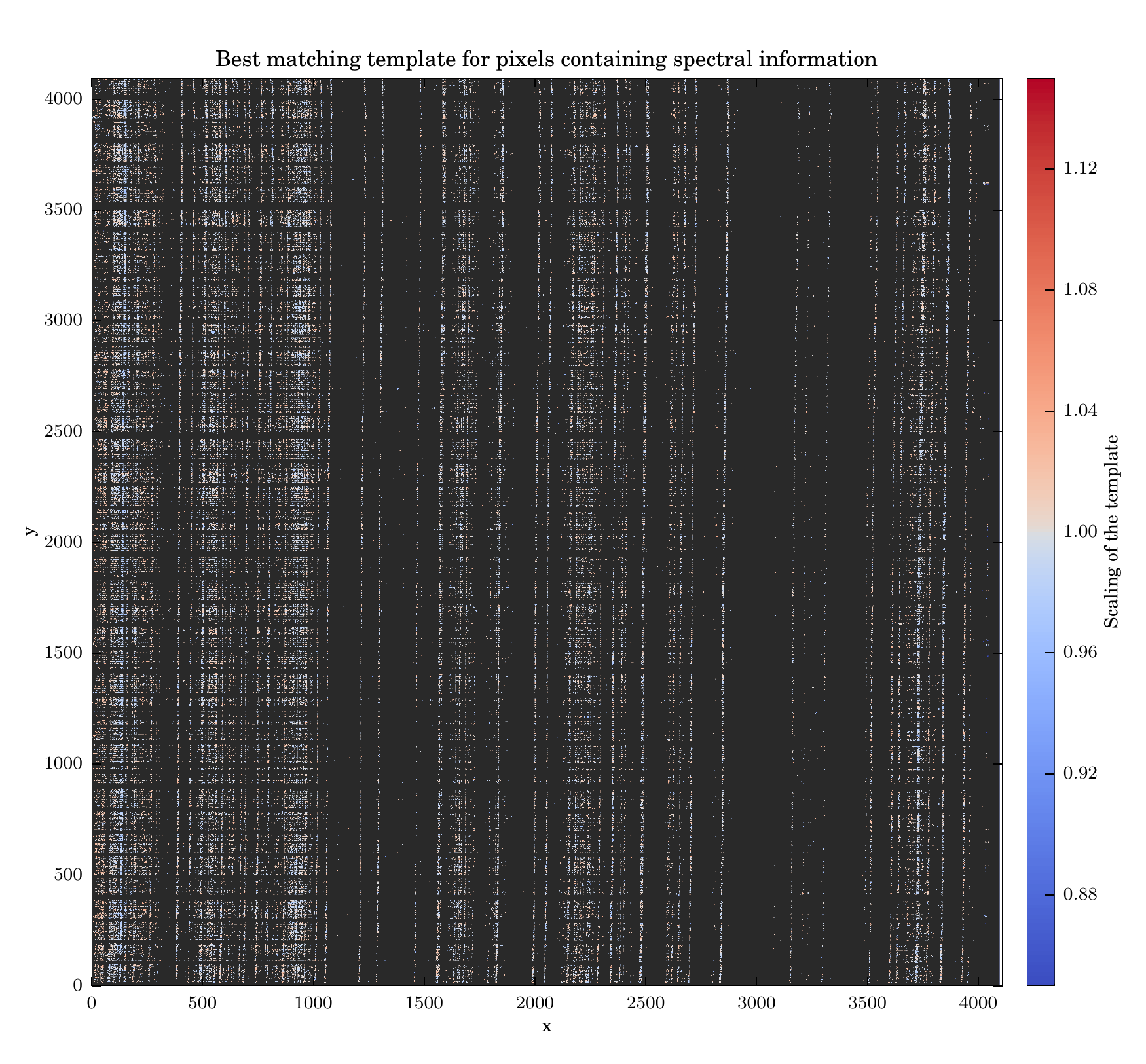}
\caption{Colour coded pixels show which template used to reconstruct the image best matches the observed image. Only a fraction of the pixels contain enough information for the residuals to be sensitive to a varying template. Pixels that do not carry enough information are coloured dark gray. The whole green arm image is shown.}
\label{fig:result1}
\end{figure*}

With a residual image made with a range of templates, we can plot for each pixel which template gives smallest residuals. Because most pixels include no or negligibly weak stellar lines, the selection of the template is irrelevant for these pixels. For most pixels, where there are no spectral features close to them, the best template is therefore chosen at random, depending on noise and numerical precision. We use a weights image -- an image of the standard deviation of residuals of all templates in each pixel -- to determine for which pixels can the best matching template be found reliably. 

\begin{figure*}
\includegraphics[width=0.95\textwidth]{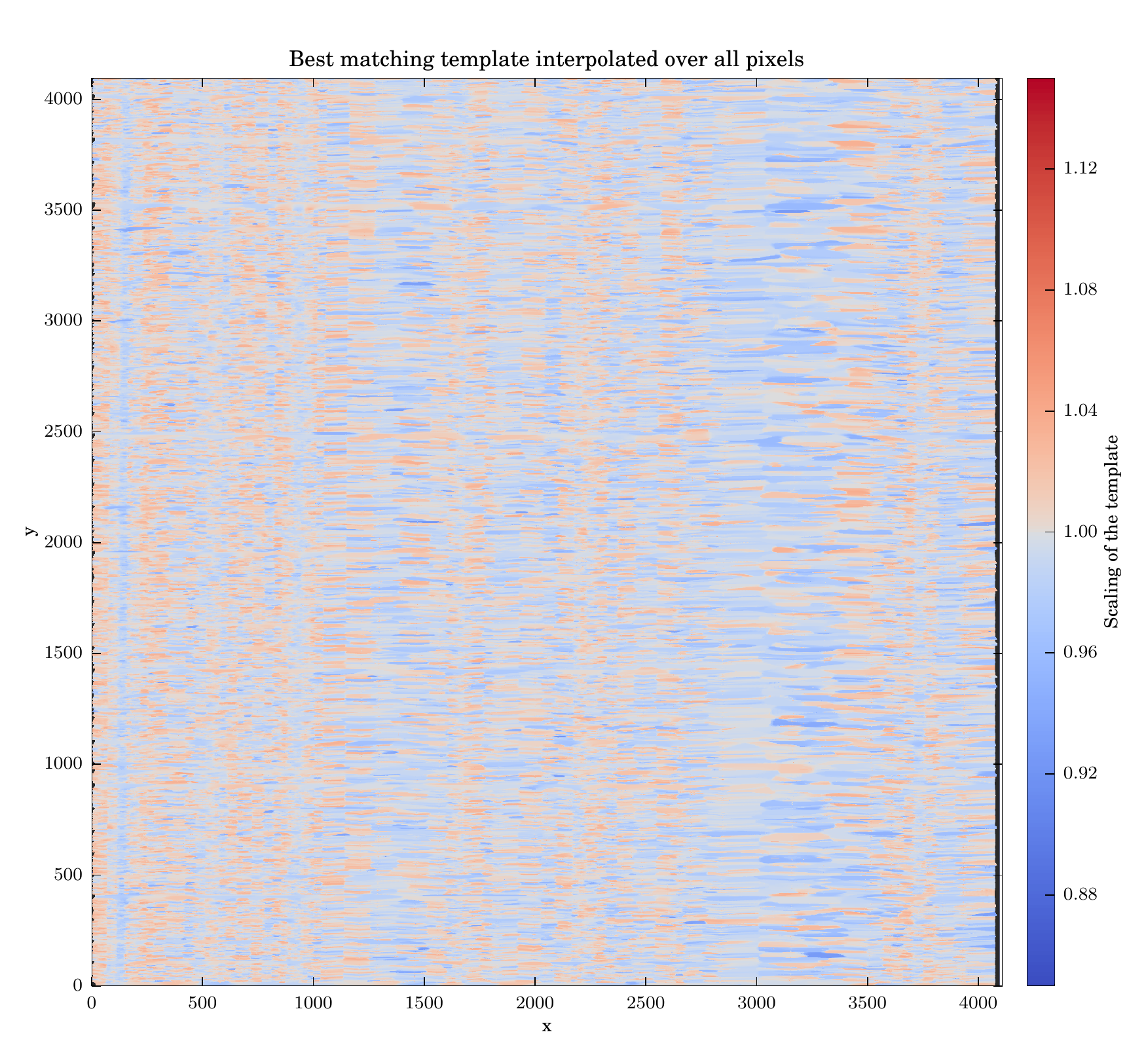}
\caption{Same as Figure \ref{fig:result1} but with gray pixels replaced with the value in the nearest valid pixel. The value from the nearest pixel ``bleeds'' predominantly into the pixels along the horizontal axis, so into the regions of the same spectral trace.}
\label{fig:result2}
\end{figure*}

\begin{figure*}
\includegraphics[width=0.95\textwidth]{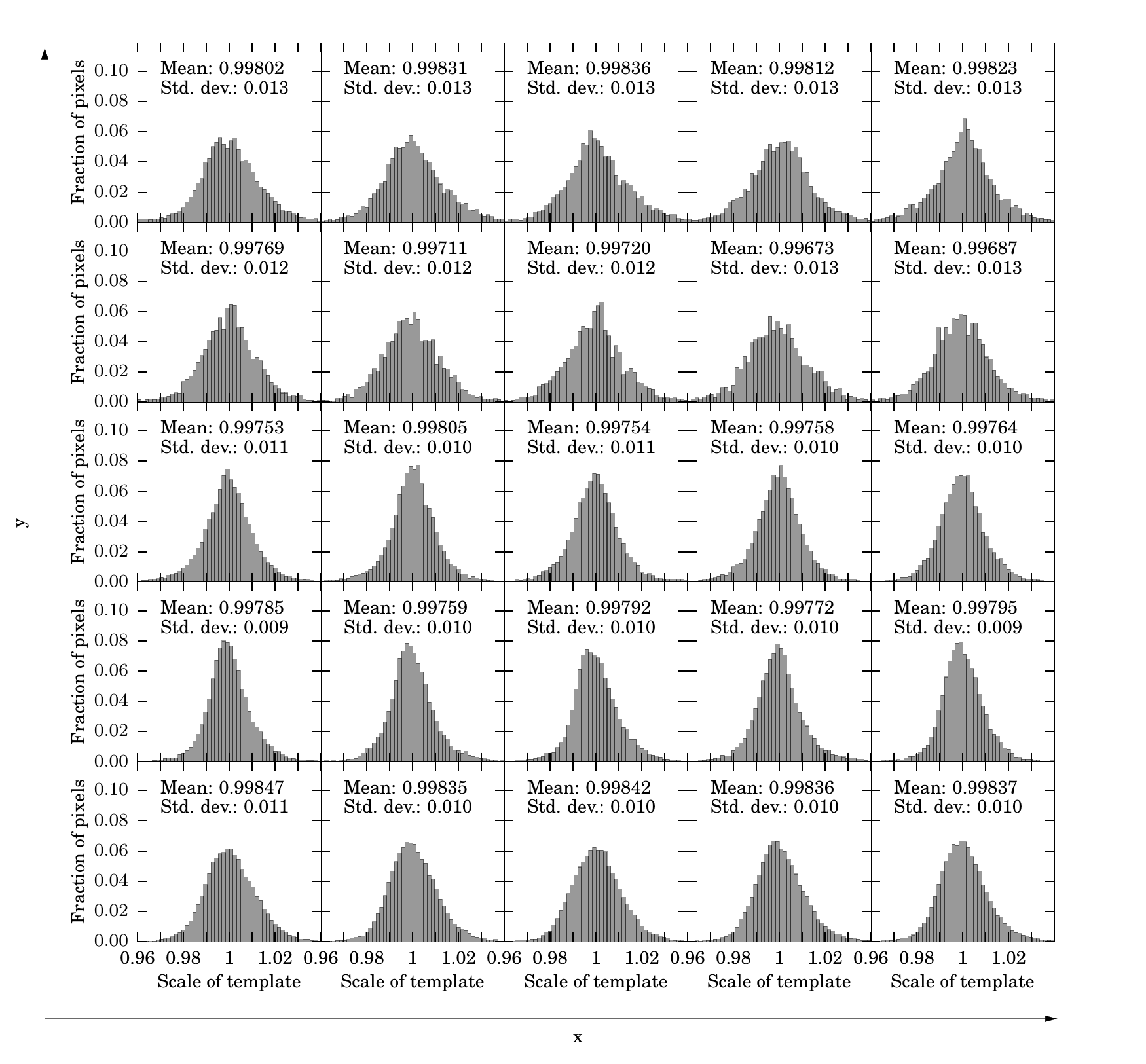}
\caption{Best templates for pixels in 25 regions of the green CCD image. Each histogram shows the statistics for a region of the image covered by that panel. Mean and standard deviation are also given explicitly in each panel.}
\label{fig:result3}
\end{figure*}

Figure \ref{fig:result1} shows the best matching template for each pixel that carries enough information. A decision of which pixels to display was made by thresholding the weights image. 

A mean value over the whole image for the scale of the best template is 0.996 with a standard deviation of around 0.01. The difference from 1.0 is probably due to a poorly determined level of scattered light. If an incorrect amount of scattered light is assumed, the relative strengths of spectral lines are affected, as the spectrum is denormalized in the same manner regardless the scattered light.

Note that Figure \ref{fig:result2} shows no pattern that matches the behavior of the optical aberrations. The same template is generally equally good in the corners -- which are most affected by the aberrations -- as well as in the middle of the image. There is no correlation between Figure \ref{fig:result2} and images in Figure \ref{fig:decomposed_green} or \ref{fig:moments_cheb_all}. The only obvious patterns are vertical bands, where a different template consistently fits better a small region in the spectrum, like at $x=150$ and around $x=3700$. Such structures could be produced by lines that are misrepresented in the template spectrum either due to natural variations in line strengths \citep{livingston82, unruh99, livingston07} or uncertainties during reduction and calibration. There is also more noise in regions where there are no or only a few very weak lines, like in two bands between $x=1100$ to $x=1500$ and $x=3000$ to $x=3500$. Here the measured value ``bleeds'' into many other pixels, as there are no other valid measurements nearby.

\subsection{Increased resolution}
Result of a traditional spectroscopic extraction is a 1D spectrum with some resolution profile. Typically the resolution varies, as it is the direct consequence of changing optical aberrations across the field. In our case the template used to reproduce a spectrum has a much higher resolution than the nominal resolution of the spectrograph. If a template with a resolution similar to nominal was used, the reproduced spectrum would have much lower resolution than intended, as the the template would be convolved twice; once to produce a lower resolution template and once to reproduce the image. 

\begin{figure}
\includegraphics[width=0.95\columnwidth]{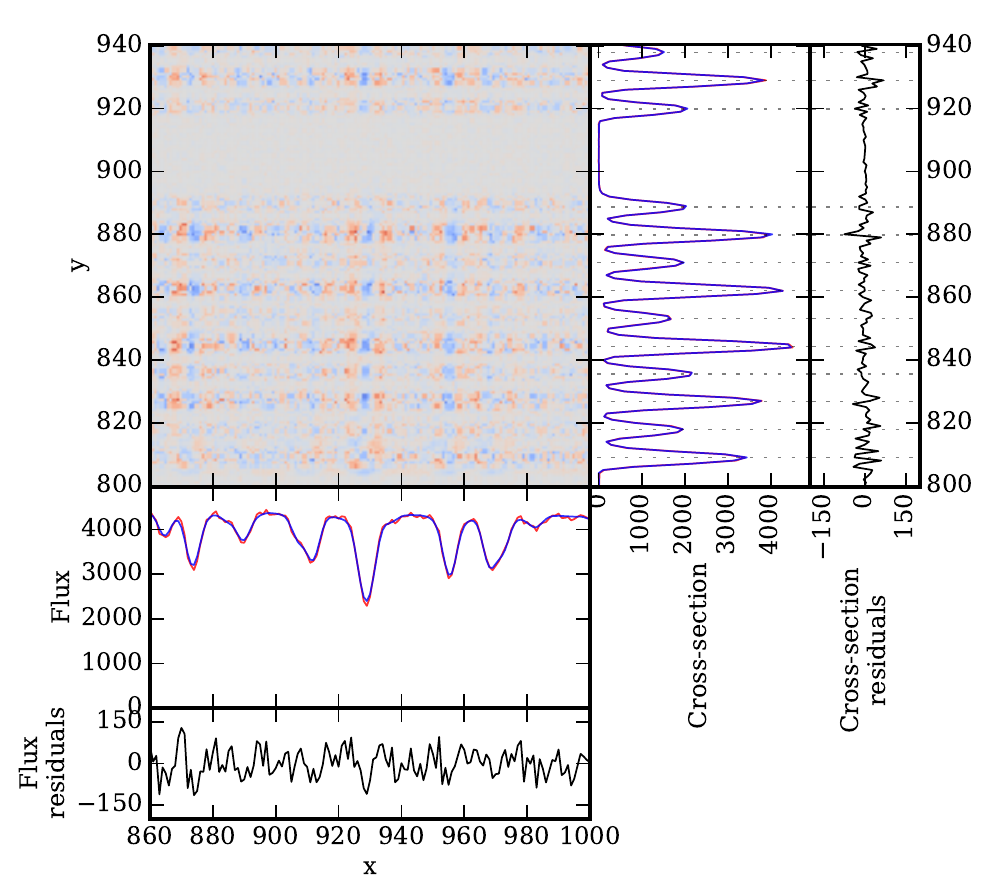}\\
\hbox{\hspace{0.47cm}\includegraphics[width=0.94\columnwidth]{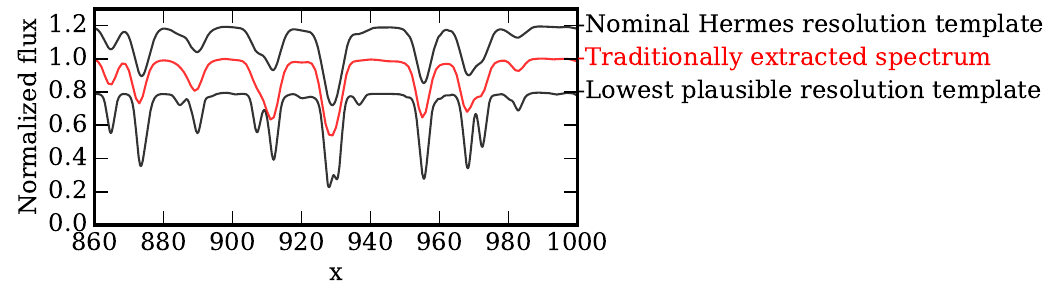}}
\caption{Top: Residuals, displayed in the same way as in Figure \ref{fig:residuals_detail}, but with a template with resolving power R=65,000 instead of R=$\infty$. Increased residuals are obvious. Bottom: Spectrum extracted in a traditional way and normalized (red) and two templates (black), one convolved to a nominal Hermes resolving power of 28,000 (top line) and one convolved to the lowest resolving power (R=65,000) we can use to reconstruct the image before residuals burst (bottom line). Templates are shifted in flux by 0.2. Note that the template with the nominal Hermes resolution and the extracted spectrum do not match very well, because the template has been convolved with a Gaussian to degrade the resolution, but not with the PSF.}
\label{fig:reproduced_res}
\end{figure}

\begin{figure}
\includegraphics[width=0.95\columnwidth]{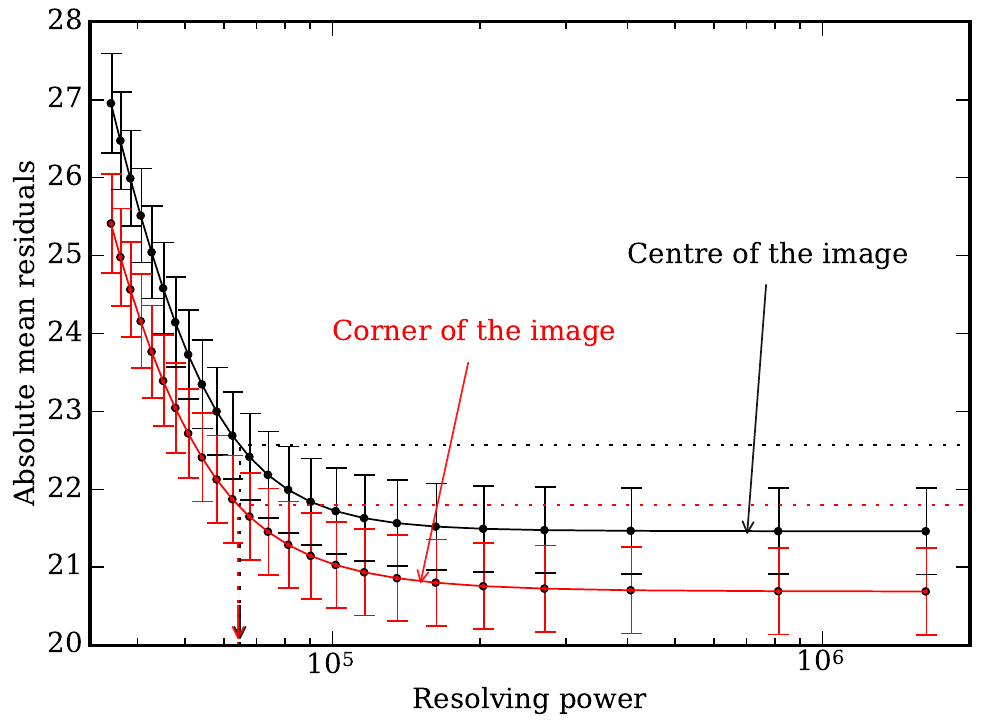}
\caption{Mean absolute residuals in the image centre (black) and one of the corners (red) as a function of Resolving power of the template used to reproduce the image. Error bars show the spread (1 sigma) of residuals after a weighted average over the whole PSF element ($15\times15$ pixels in size). We claim that the difference between a template with infinite resolution and a template with finite resolution is significant at the two sigma level of the spread at large resolving powers. This corresponds to a template with resolving power of around 65,000, as indicated by the dashed lines. This value is almost the same all over the image.}
\label{fig:degraded}
\end{figure}

The image was reproduced with a range of templates with different resolutions. One such reproduction is displayed in Figure \ref{fig:reproduced_res}. Compared with Figure \ref{fig:residuals_detail}, the residuals increase significantly when the degraded template is used. To find out how degraded the template can be before the differences become noticeable, we plot the mean residuals as a function of the resolving power of the template (Figure \ref{fig:degraded}). At very high resolving powers no variations in the residuals are detected. When the template is degraded more, the residuals start to rise slowly. Same trend is observed in the centre and in the corners of the image. The only difference is that different regions have different mean residuals due to different flux in that part of the image. The scatter of residuals from pixel to pixel (illustrated by the error bars in Figure \ref{fig:degraded}) is mostly the same everywhere. We can comfortably say that the reproduced spectrum is noticeably different when a template with R=65,000 or lower is used. 

After successfully reproducing the image, one can proclaim the best matching template to be the inferred or extracted 1D spectrum. Since the difference between templates with very high resolutions is negligible, the lowest resolution template where the difference becomes significant should be used as the extracted spectrum. A comparison of a traditionally extracted 1D spectrum and the best matching template is shown in Figure \ref{fig:reproduced_res} in the bottom panel. The resulting template shows more detail than the extracted spectrum. Structures significantly smaller than the PSF are resolved. Theoretically, a traditionally extracted spectrum can be further analyzed to reveal finer structures, but this might be a very challenging task. The strongest line in the middle of the plot, for example would hardly be suspected to have two components. Even the line at $x=970$ would be hard to fit with two components, but in the best matching template the two lines are well separated. With our schema a complicated problem is very simplified, as degeneracies between blended lines and impact of sometimes poorly known LSF is eliminated. One does not have to deconvolve the traditionally extracted spectra or deblend the lines, if the best matching template is used. Any uncertainty in the contribution of individual components of deblended lines is conveniently translated into error bars on the flux every pixel of the extracted 1D spectrum would have. With only one Solar template at our disposal we can not produce these error bars, as there is no practical and realistic way to perturb individual lines in the template. But in a different application where a range of templates is available, the error bars would come naturally from a large number of templates that would have to be tried before the best match is found.

\section{Discussion}

With current multi-fibre spectrographs we are close to the limit where larger instruments using more fibres and covering larger field of view cannot be built without excessively big and expensive optics, as hinted by GALAH \citep{sheinis15} or WEAVE \citep{dalton14}, or splitting the fibres between several instruments like in the LAMOST survey \citep{cui12}. Existing instruments were build with limitations of a traditional reduction in mind. Our approach allows one to reduce the tolerances of the spectrograph design and increase the rigorosity of the reduction process, though at a cost of more complicated analysis. However, merging some analysis with the reduction makes other steps easier, like removal of telluric absorptions. Tellurics as weak as those shown in Figure \ref{fig:residuals_tel}, for example, are not removed during the reduction in the GALAH survey, as their signal is lost in the stellar spectral lines.

The main result of this work is an algorithm presented in Section \ref{sec:section4} describing a practical way of a complete reconstruction of a multi-fibre spectral image affected by any kind of optical aberrations. The PSF can be alternatively measured by other means than a photonic comb and the template spectra can be introduced in a number of ways. We chose to demonstrate the method on Solar spectra using a standard Solar spectrum as a template, but discuss other more general possibilities below.

While the reproduction process is the same for Solar and stellar spectra, finding the correct templates for the stellar spectra is much harder. Unlike high resolution observed Solar spectrum that we used here, large collections of stellar templates can only be synthetic and suffer from missing spectral lines and errors in the linelists. In recent years algorithms for a rapid production of synthetic spectra have been developed \citep{ting16, rix16}, but the available linelists are only reliable for a relatively small number of lines that are frequently studied. The linelists can only be significantly improved by observations and our method is very suitable for that, as it deals very carefully with all possible systematic errors arising from the optical aberrations.

Within our schema a 1D spectrum can only be extracted if a template is available. This does not mean that a synthetic template must be used. An empirical template can be constructed iteratively. After the image is reconstructed with an approximate template, the template can be varied until the residuals are minimized.  This is a slow process, but allows us to extract any spectrum, regardless the availability of synthetic templates. Despite many iterations that are needed to iteratively extract a 1D spectrum, we avoid all big matrix inversions, like proposed in \citet{bolton10}, which makes extraction process faster and easy to parallelize. Our approach can also be used to improve stellar templates. Ability to extract spectra way above the nominal resolution of the instrument is a further advantage, if linelists are to be improved. In the GALAH survey some observations are systematically repeated, so the stellar templates produced in the above way could be quickly verified by comparing spectra produced with a different PSF and a different fibre from any of the two sets. 

In the GALAH survey we are exploring practical solutions for extraction of 1D spectra with the above scheme and plan to implement it in the reduction pipelines. Current synthetic templates proved unsuitable for the precision we require even for the derivation of the most basic stellar parameters. We are focusing into a general extraction of 1D spectra where synthetic templates only serve as a first approximation. The reason that the 1D extracted spectra are wanted as opposed to merging the analysis and reduction, as envisioned in this work, is that all present analysis methods are designed to use 1D spectra. It would be unwise to invest too much time into merging the analysis into the reduction while neglecting the proven methods and delaying the scientific outputs. Data-driven astronomy is also becoming increasingly more popular \citep{ness18}, which in the case of stellar spectroscopy exclusively relies on 1D spectra. Therefore being able to provide 1D extracted spectra is a priority for any reduction technique.

\section*{Acknowledgements}
JK is supported by a Discovery Project grant from the Australian Research Council (DP150104667) awarded to J. Bland-Hawthorn and T. Bedding. SLM acknowledges funding from the Australian Research Council through grant DP180101791. TZ acknowledge the financial support from the Slovenian Research Agency (research core funding No. P1-0188). DMN was supported by the Allan C. and Dorothy H. Davis Fellowship. A. R. C. acknowledges support from the Australian Research Council through Discovery Project grant DP160100637.

\bibliographystyle{mnras}
\bibliography{bib}


\clearpage
\appendix

\section{Discrete Chebyshev moments}
\label{sec:app_all}

Figures in this appendix show all discrete Chebyshev moments for all arms and both fibre sets. Measured moments, decomposition into a smooth and fibre model, and the residuals are shown.

\begin{figure*}
\includegraphics[width=0.95\textwidth]{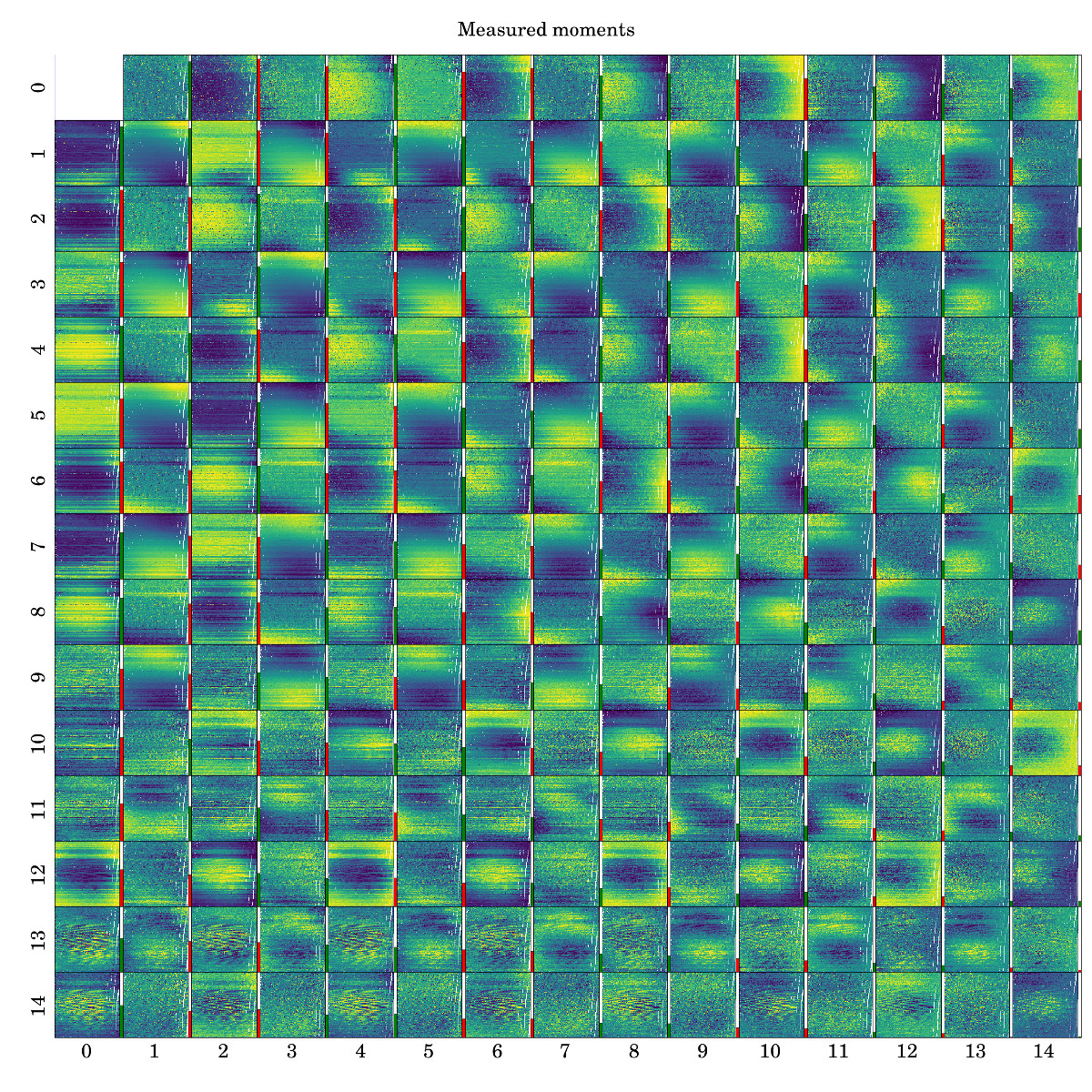}
\caption{Measured power of discrete Chebyshev moments for green arm and fibres from plate 0. Each small panel shows a different moment, marked with the numbers on the left and bottom. Each measured photonic comb peak is displayed with a point and the power of the moment is colour coded. Colour ranges are normalized differently in each panel, because the powers of different moments span $\sim30$ orders of magnitude. A narrow bar on the right of each panel shows the power of that moment averaged over the whole CCD plane. The higher the bar, the stronger the moment. Scale is logarithmic and the whole range extends over 30 orders of magnitude. Green bar means that moment has a positive value and red means a negative average value. Moment 0,0 is not plotted. It is set to 1 for the whole CCD plane, so the total flux of each PSF is normalized to 1.}
\label{fig:moments_cheb_all}
\end{figure*}

\begin{figure*}
\includegraphics[width=0.95\textwidth]{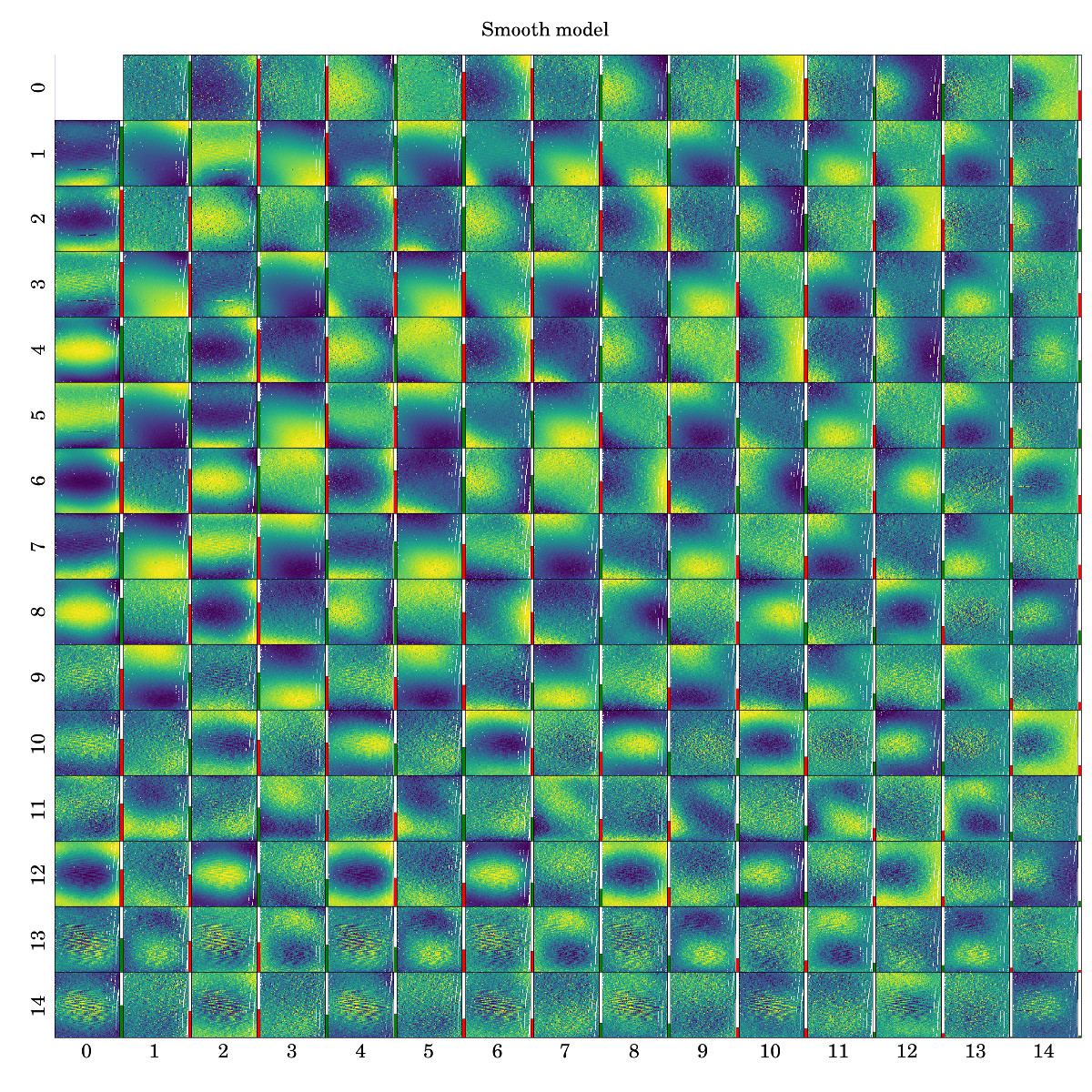}
\caption{Smooth part of the discrete Chebyshev moments for green arm and fibres from plate 0.}
\end{figure*}

\begin{figure*}
\includegraphics[width=0.95\textwidth]{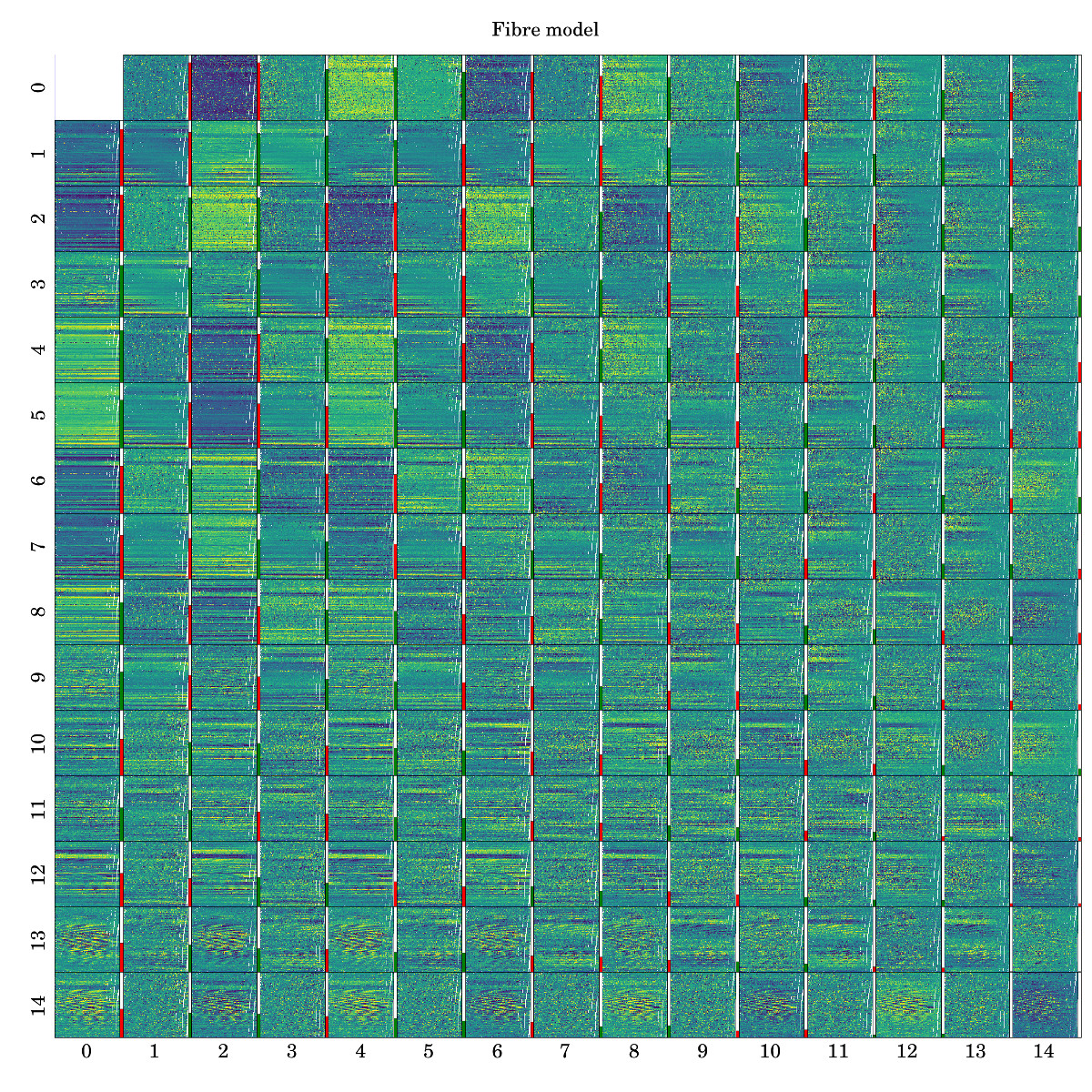}
\caption{Fibre part of the discrete Chebyshev moments for green arm and fibres from plate 0.}
\end{figure*}

\begin{figure*}
\includegraphics[width=0.95\textwidth]{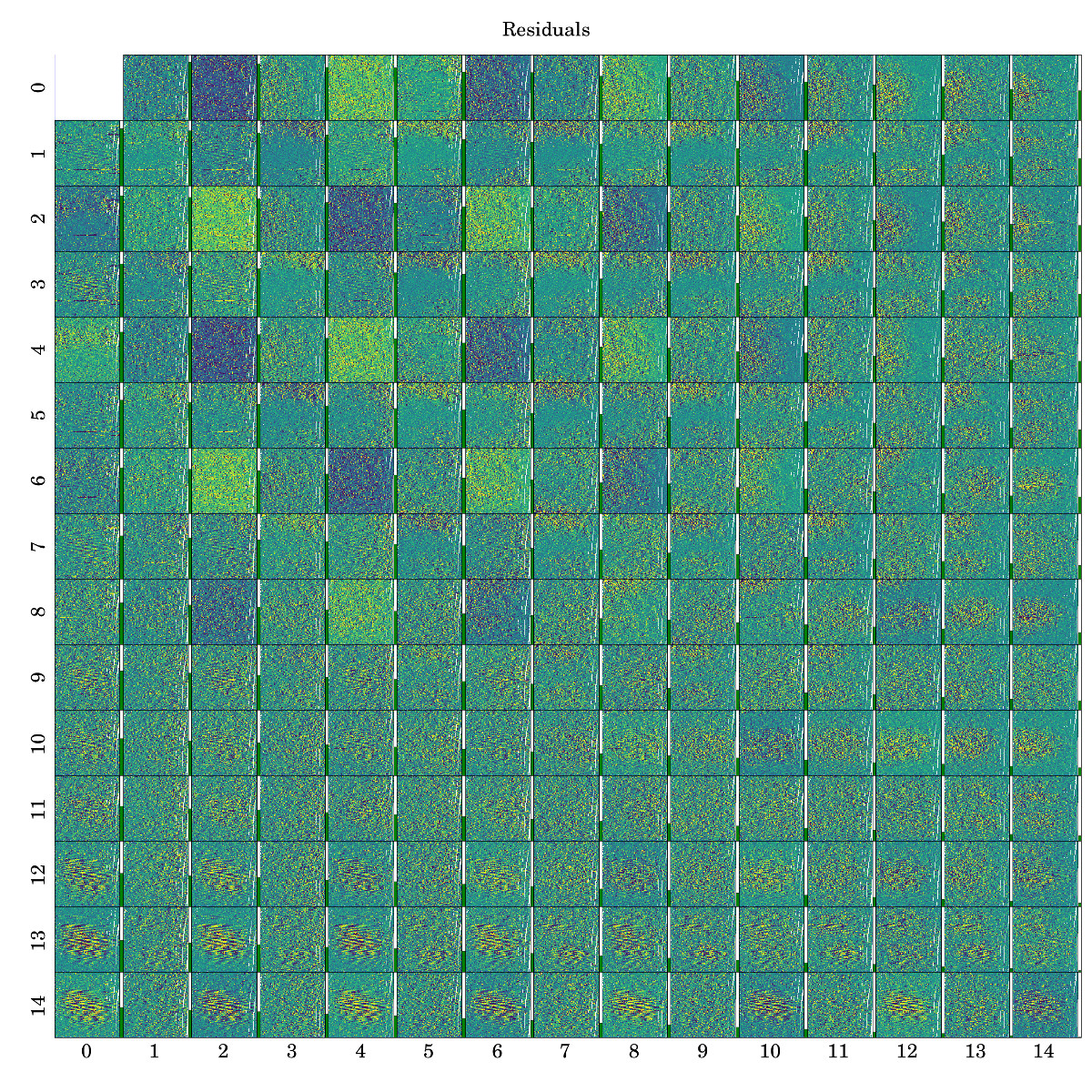}
\caption{Residuals between the modeled and measured discrete Chebyshev moments for green arm and fibres from plate 0. Bars on the right of each  panel show the average standard deviation of the residuals in that panel.}
\end{figure*}

\bsp
\label{lastpage}
\end{document}